\documentclass[manuscript,screen]{acmart}

\AtBeginDocument{%
  \providecommand\BibTeX{{%
    \normalfont B\kern-0.5em{\scshape i\kern-0.25em b}\kern-0.8em\TeX}}}

\setcopyright{acmcopyright}
\copyrightyear{2018}
\acmYear{2018}
\acmDOI{10.1145/1122445.1122456}

\acmConference[Woodstock '18]{Woodstock '18: ACM Symposium on Neural
  Gaze Detection}{June 03--05, 2018}{Woodstock, NY}
\acmBooktitle{Woodstock '18: ACM Symposium on Neural Gaze Detection,
  June 03--05, 2018, Woodstock, NY}
\acmPrice{15.00}
\acmISBN{978-1-4503-XXXX-X/18/06}

\usepackage{graphicx}
\usepackage{epstopdf}
\usepackage{epsfig}
\usepackage{ulem}
\usepackage{amsmath}
\usepackage{algorithm}
\usepackage{algorithmic}
\usepackage{subcaption}
\usepackage{multirow}
\usepackage{amsmath}%
\usepackage{tabularx}
\usepackage{booktabs}
\usepackage{graphicx}

\newcommand{\tabincell}[2]{\begin{tabular}{@{}#1@{}}#2\end{tabular}}

\begin{document}

\title{GUX-Analyzer: A Deep Multi-modal Analyzer Via Motivational Flow For Game User Experience}

\author{Zhitao Liu}
\email{zl425uestc@gmail.com}
\orcid{0000-0003-3499-4157}

\affiliation{%
  \institution{School of Aeronautics And Astronautics, University of Electronic Science and
Technology of China}
  \city{Chengdu}
  \country{China}
  \postcode{611731}
}

\author{Ning Xie}
\authornote{Corresponding authors}
\affiliation{%
  \institution{Center for Future Media and School of Computer Science
and Engineering, University of Electronic Science and
Technology of China}
  \city{Chengdu}
  \country{China}}
  \postcode{611731}
\email{seanxiening@gmail.com}
\orcid{0000-0002-1509-464X}

\author{Guobiao Yang}
\affiliation{%
  \institution{Center for Future Media and School of Computer Science
and Engineering, University of Electronic Science and
Technology of China}
  \city{Chengdu}
  \country{China}}
  \postcode{611731}
  \email{923064497@qq.com}
  \orcid{0000-0002-8121-5315}
  
\author{Jiale Dou}
\affiliation{%
  \institution{Center for Future Media and School of Computer Science
and Engineering, University of Electronic Science and
Technology of China}
  \city{Chengdu}
  \country{China}}
  \postcode{611731}
  \email{834105473@qq.com}
  \orcid{0000-0002-5285-5196}
  
  \author{Lanxiao Huang}
\affiliation{%
 \institution{Timi L1 Studio of Tencent Corp, Tencent}
   \city{Chengdu}
  \country{China}}
  \email{jackiehuang@tencent.com}

\author{Guang Yang}
\affiliation{%
  \institution{Timi L1 Studio of Tencent Corp, Tencent}
   \city{Chengdu}
  \country{China}}
  \email{mikoyang@tencent.com}

\author{Lin Yuan}
\affiliation{%
  \institution{Timi L1 Studio of Tencent Corp, Tencent}
   \city{Chengdu}
  \country{China}}
  \email{tayloryuan@tencent.com}
  
\renewcommand{\shortauthors}{Zhitao Liu, et al.}

\begin{abstract}
Quantitative analysis of Game User eXperience (GUX) is important to the game industry. 
Different from the typical questionnaire analysis, this paper focuses on the computational analysis of GUX. 
We aim to analyze the relationship between game and players using the multi-modal data including physiological data and game process data. 
We theoretically extend the Flow model from the classic skill-and-challenge plane by expanding new dimension on motivation, which is the result of the multi-modal data analysis on affect, and physiological data. 
We call this 3D Flow as Motivational Flow, MovFlow. 
Meanwhile, we implement a quantitative GUX Analysis System (GUXAS), which can predict the player’s in-game experience state by only using game process data. 
It analyzes the correlation among not only in-game state, but the player’s psychological-and-physiological reaction in the entire interactive game-play process. 
The experiments demonstrated our MovFlow model efficiently distinguished the users’ in-game experience states from the perspective of GUX. 

\end{abstract}



\maketitle

\section{Introduction}

\begin{figure}
  \includegraphics[width=\textwidth]{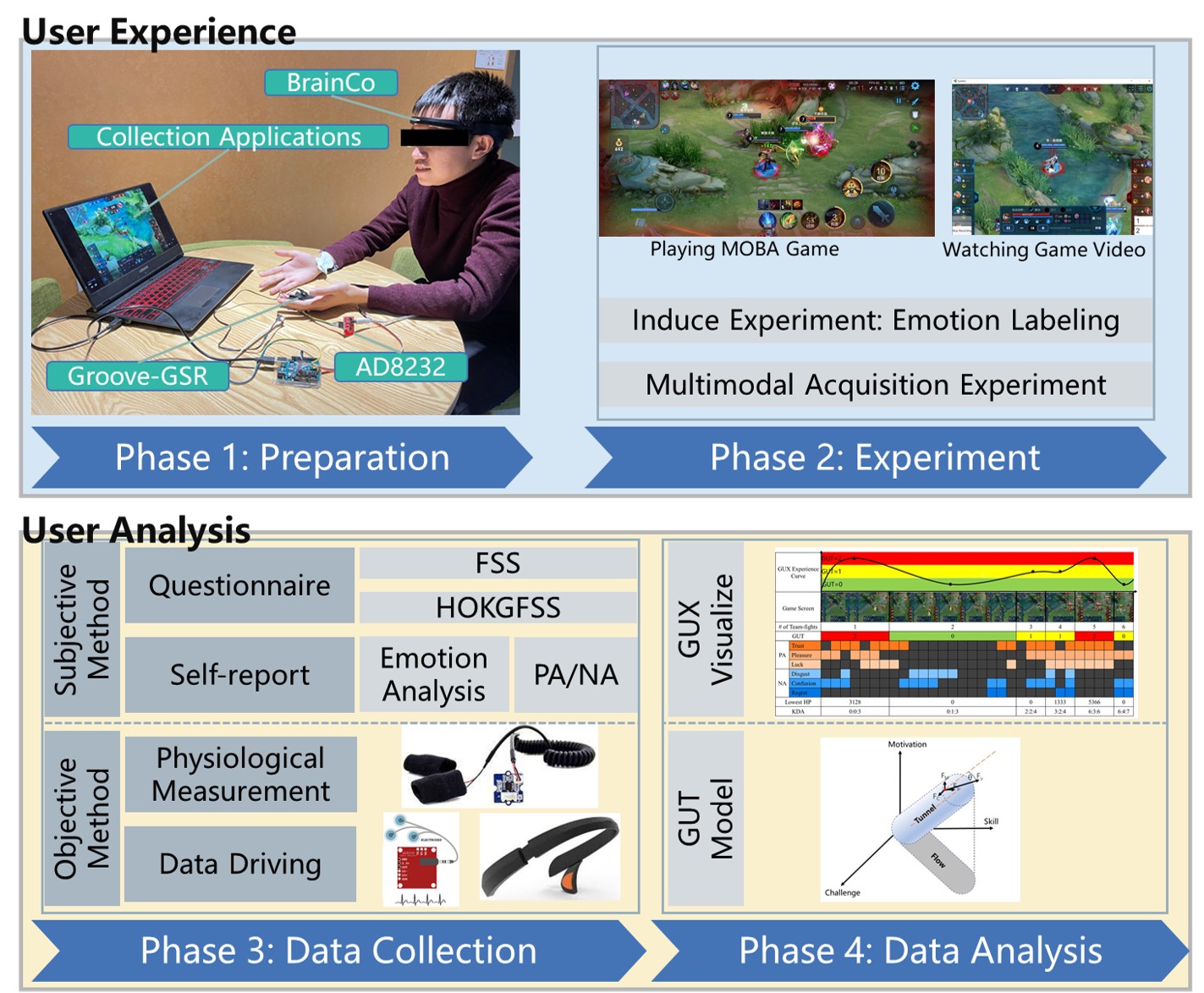}
  \caption{Proposed pipeline. Phase 1 shows a user wearing our proposed self-made wearable device to conduct the task. Phase 2 shows the scene of our experiment. Phase 3 \& 4 illustrate the collection and the analysis of data. The participant needs to wear physiological devices when conducting the task. By collecting and analyzing the physiological data and game process data (game log) of the experiment, this paper aims to expand the existing work and study a more general GUX detection model using the physiological data collected by the wearable device and game play data.}
  \Description{Enjoying the baseball game from the third-base
  seats. Ichiro Suzuki preparing to bat.}
  \label{fig:teaser}
\end{figure}

Game User eXperience (GUX) is understood as the subjective relationship between users and games, which refers to the perception and the user's responses during the game-play~\cite{bernhaupt2010evaluating}.
Flow analysis is an appealing tool in GUX research used to evaluate whether users are just in the specific state called Flow ~\cite{csikszentmihalyi1990flow}. That is, whether the user's skill can match the challenge of the task perfectly. Different from analyzing the subjective questionnaire after the game-play, the mainstream methodology nowadays is in the real-time manner, in which researchers collect the users' feedback and analyze the data without interrupting the game-play in the online manner. The typical Flow analysis focuses on obtaining the users' status by comprehensive considering the users' skills and the challenges of the tasks in these two aspects. The scholars applied the Flow analysis as the appealing measurement tool \cite{ye2020flow,chanel2011emotion,DBLP:conf/ijcai/MaierEMZF19,10.1145/3170427.3188480,berta2013electroencephalogram}. It is called {\it 2D Flow} from the aspect of the dimension of the data. But, the 2D Flow still has limitation in practical data-driven manner, since it ignores the users' motivation. 

In GUX, the motivation is one of the most powerful motivating factors for people. Many scholars believed that user experience is the result of behavior inspired by a certain motivation in the specific environment~\cite{makela2001supporting}. In the game-play, the competitive situation itself can induce strong motivation. The motivation indicates how much the player seek to satisfy psychological needs in the context of game-play\cite{ryan2006motivational}. Context can be understood as any or all information that characterizes the situation of a certain entity, which could be a person, place, or object relevant for user-product interaction \cite{dey2001understanding}. During that, the players enable to generate numerous rich affects so as to motivate them to keep forging ahead to meet the challenge~\cite{vandercammen2014relating}. It has been demonstrated that people who experience positive affects are more motivated for a pleasant task~\cite{isen2005influence}, experience more interest and enjoyment while carrying out the task~\cite{erez2002influence,isen1987positive,staw1993affect}, and continue to work longer on less pleasant, and even uninteresting tasks~\cite{isen2005influence}. Considering the Flow and the motivation, ideally, when users' skills and challenges tend to be balanced, under the influence of the users’ motivation, positive experiences will be activated more than negative ones, and users will be more likely to develop positive affects. Specifically, the positive experience is defined as anything that increases users' pleasure, immersion, and/or the challenge of the game, whereas the negative experience is defined as any situation where the player is bored, frustrated or want to quit the game~\cite{Desurvire2015}.

In our work, we carefully study on the concepts of the Flow and the motivation. We propose a new 3D computational GUX model called the {\it 3D MovFlow}, where we expand the new dimension of motivation on the typical 2D flow~\cite{csikszentmihalyi1990flow}. We firstly describe the 3D MovFlow space mathematically. As illustrated in Figure 2, the positive experience is constrained to the inside of the tunnel, named as {\it GUX Tunnel, GUT}. In order to further guarantee that the GUT facilitates the classification of the user experience in practice, we choose the data-driven methodology to refine the surface of the GUT. Our aim is to quantitatively analyze the user's experience using the multi-model data not only the users' response such as physiological and psychological signal, but also the in-game process logs. For the motivation of computing, we propose the method among the multi-modal physiological and psychological data including the electroencephalogram (EEG), the electrocardiogram (ECG), the Galvanic Skin Response (GSR), the attention, the meditation, and the affect. To collect these data, we design and implement a compact wearable multi-modal physiological and psychological data acquisition system to collect user’ data during the game match. Then, we propose the GUX analysis system for the user's state prediction, called {\it GUXAS} as illustrated in Figure 3. During the training phase, we will analyze the players' GUX state via game process data and physiological data (collected by the GUXAS Multi-modal \\Data Acquisition system as illustrated in Figure 5), and analyze the state only through game process data during the in practical phase. We will than visualize the users' GUX state and game process data for game designers re-designing the game as illustrated in Figure 4.
In order to classify the GUX state, we proposed a deep sequential correlation model based on the Dynamic Time Warping (DTW) matching algorithm. \\
Then, we combine the few shot metric learning based on the Siamese network and personality embedding together to predict the user’s GUX state. 
We setup two experiment to acquisition user's physiological data and game process data, and verify the 3D MovFlow model and the GUXAS system we proposed.   
The experiment demonstrated that the 3D MovFlow can efficiently inference the user's state of GUX from the game process. The GUXAS system ultimately facilitates the prediction of the player's in-game user experience state by using only their game process data.
We believe that GUXAS system and 3D MovFlow model is an efficient tool to improve the game design from the perspective of user experience, especially to the games with complex battle arena such as Multiplayer Online Battle Arena (MOBA).


\begin{figure}
  \centering
  \begin{center}
\includegraphics[width= 0.5\columnwidth]{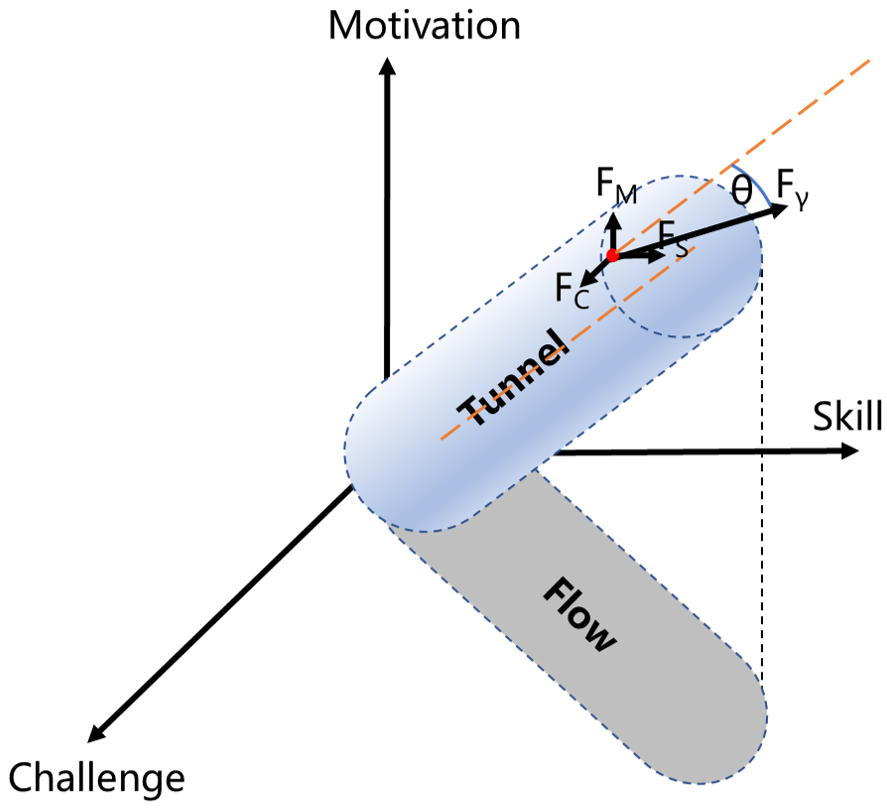}
  \caption{Diagram of the proposed 3D MovFlow (GUT) theoretical model (The state when the player's motivation dimension has not reached its peak), where Z-axis is Motivation, Y-axis is Skill, and X-axis is Challenge. Location inside the tunnel indicates a good GUX state. The GUX state located on the orange dash line is the best GUX state. The equation of the orange line is X (Challenge)=Y (Skill)=Z (Motivation). The projection of the tunnel on the Skill and Challenge planes is Flow. We can judge the player's GUX state by observing the direction of the movement trend (the direction of $F_\gamma$) generated by the combined forces of the three dimensions in space at a certain moment.}
    \end{center}
\end{figure}

\begin{figure*}
\centering
\includegraphics[width= \textwidth]{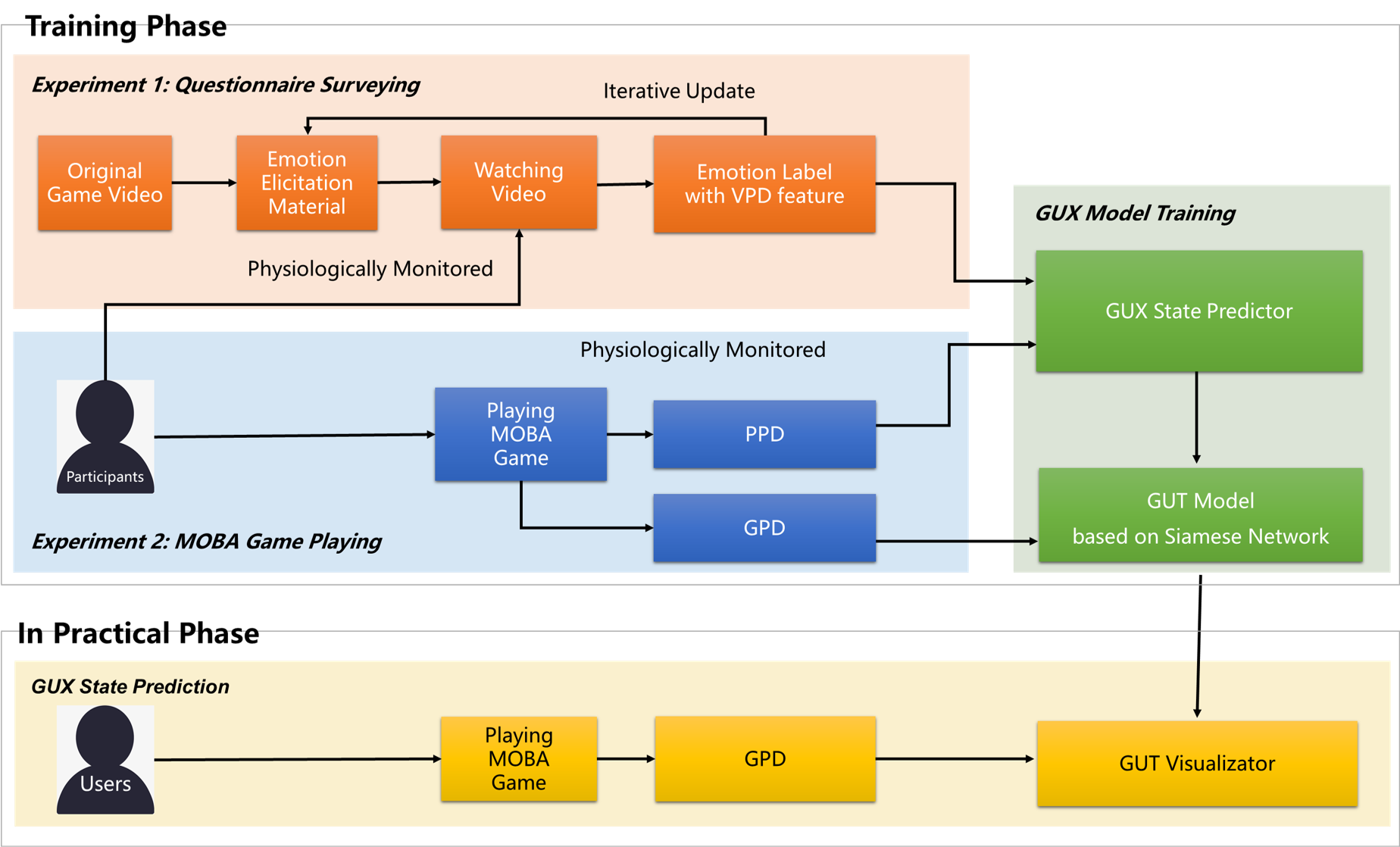}
\caption{Schematic diagram of GUXAS pipeline. The orange area is experiment 1, where the participants need to carry our self-made lightweight wearable device, and watch the affect eliciting MOBA game video. Also, the participants need to confirm the affect label of VPD (the physiological data of the subject when watching the video) through a semi-structured interview.
The light blue area is experiment 2, where participants are required to play a MOBA game, and then we collect the Play-time Physiological Data (PPD) and game process data of the player during the game-play. Using the Dynamic Time Warping (DTW) algorithm, we predict the GUX state of participants with the VPD features and labels. The result used as input to the Siamese network to train the GUX state label (green area) together with the game process data. 
We finally achieve inputting the game process data of any player, and then output his corresponding GUX state (yellow area) during the in-game time.
}
\end{figure*}

\begin{figure*}
  \centering
 \includegraphics[width = 1\textwidth]{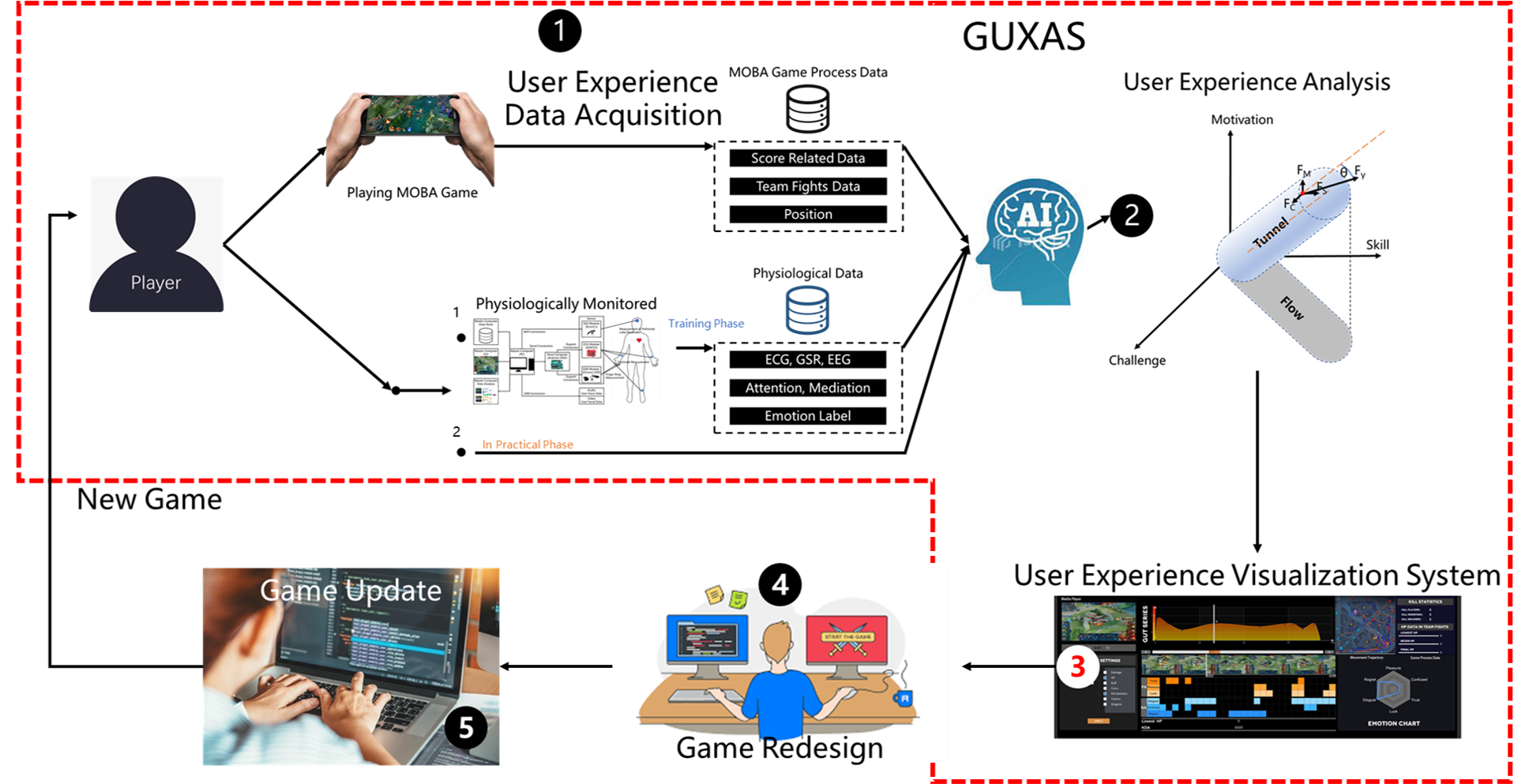}
  \caption{
GUXAS system flow chart. The Red box is our GUXAS system, which include user experience data acquisition part, user experience analysis part, and user experience visualization part. During the training phase of the GUXAS system, we need to collect physiological data from some of the users to train the AI predicting part. During the in practical phase, we will only use the game process data analyzing the user GUX state. The game will be redesigned and upgraded taking into account the feedback from our GUXAS system by game designer.  
}
\end{figure*}

\section{Related works}
Considering the importance of GUX analyses which helps game designers to improve video games experience and interaction, several works have made attempt in this domain. Hence, we discuss some related works as follows. Mihaly Csikszentmihalyi~\cite{csikszentmihalyi1990flow} proposed the concept of Flow, which is defined as a highly focused mental state conducive to productivity. Researchers have proven that Flow detection can be based on physiological data including EMG \cite{kivikangas2006psychophysiology,de2010psychophysiology,2010Affective}, Meditation \cite{csikszentmihalyi1990flow,2010Effortless}, Cortisol \cite{baumann2010seeing}, Cardiovascular measures \cite{keller2011physiological,de2010psychophysiology} and EDA \cite{kivikangas2006psychophysiology}. These methods quantitatively measure the users' Flow during the game, including context \cite{ENGL201383}. Due to the difficulty of collecting the EMG, and Cortisol measurement, our experiment mainly used ECG (Cardiovascular measures), GSR and EEG (Meditation) as the main physiological data dimensions to measure Flow. 

The work of Mandryk \cite{mandryk2007fuzzy} opens the possibility of measuring GUX using physiological data. Since then, many researchers such as \cite{giakoumis2011automatic} have begun to study the relationship between GUX, cognition and physiological data (such as ECG and GSR signals) to identify users' boring experience when playing video games.
This work \cite{nacke2008flow} uses an FPS game called "Half-Life 2" to measure users' Immersion and Flow state during the game play. This study is worth mentioning because it is the first work that changes the user's game experience by modifying the elements in the game. However, this work only used statistical methods to analyze the differences in the physiological data generated by the different game elements but did not conduct further multi-modal sequence quantitative analysis.

Using classic two-dimensional experience model to predict the Flow state of users during game play, researchers  ~\cite{ye2020flow,chanel2011emotion,DBLP:conf/ijcai/MaierEMZF19,10.1145/3170427.3188480}, dynamically adjust the challenges in the game while ensuring that the users' skill remains basically unchan-\\-ged. They aimed at finding the characteristics of physiological data corresponding to the different Flow states. 
They collected multiple physiological data including GSR, HR, RSP, and EEG signals from participants.
However, the game experience of players analyzed in this way is still quite different from the GUX in the real complex environment.
The previous research mainly has the following limitations. They are mainly limited to small games and single data dimensions. 
They neglect the impact of the game itself and the effect from the player's motivation on user experience.

Motivation directly affect the user’s GUX state during the in-game time \cite{hassenzahl2006user}, and some researchers even think that user experience is the result of behavior inspired by a certain motivation in the specific environment \cite{makela2001supporting}.
Keller developed a model that includes four main strategies to elicit and maintain motivation: attention, relevance, confidence, and satisfaction (ARCS) \cite{keller2009motivational}.
At the same time, affects can generate motivation. It has been demonstrated that people who experience positive affects are more motivated for a pleasant task \cite{isen2005influence}, experience more interest and enjoyment while carrying out the task \cite{erez2002influence,isen1987positive,staw1993affect}, and continue to work longer on less pleasant, and even uninteresting tasks \cite{isen2005influence}.

In view of the shortcomings of the previous researches, this work aims to expand the existing work and study a more general GUX detection model using the physiological data collected with the self-made wearable device and game process data.

\section{Method}
In this section, we will introduce our proposed GUX state prediction system (GUXAS)  and 3D MovFlow (GUT) model, which includes four subsections, namely the data acquisition system, GUX-Tunnel theoretical model, user GUX state classification model based on DTW deep learning method and user GUX prediction model based on few shot metric learning. 
Following is the detail of our proposed method.

\subsection{3D Motivation Flow Model with Skill-Challenge-Motivation}

\textbf{2D Flow}
Classic Flow, as a mental state, was first proposed by Csikszentmihalyi, in which user will fully immersed in his (her) work and enjoy himself doing the job. According to Csikszentmihalyi's Flow theory \cite{csikszentmihalyi1990flow,csikszentmihalyi1992optimal}, Flow usually occurs when the skills of the activity participant can meet the challenging demands of the task being performed, and this balance allows the participant to better master the task and become fully engaged in it. Conversely, an unbalanced relationship between the participant's ability and the difficulty of the task can prevent the participant from entering the state of Flow and even produce boredom or anxiety. In game user experience, Flow can be understood as an extreme experience gained when the user is fully engaged in playing the game. This experience is generally accompanied by a high level of concentration, a strong sense of immersion, motivation, and excitement. In essence, Flow is characterized by the complete absorption in what one does, and a resulting transformation in one's sense of time.

However, only studying the user's Flow state, and only analyzing the user's skills and challenges to get the user's GUX state, still cannot fully understand the user's real feelings during the in-game time. 
Especially for highly competitive games, users will be very concerned about the outcome of the game and the performance of teammates and opponents during the game.
It is difficult to classify the user's GUX state comprehensively by only using the classic 2D Flow model with skill-and-challenge. 
The 3D motivational Flow model (GUT) composed of the motivation (including Affect, Attention, Meditation) and the classic flow model can provide a more comprehensive method to analysis user's GUX state.  

For the game event ‘Free Win’ (Free Win, the user does not act but can win by relying on his teammates), the user may not be in the Flow state due to skills and challenges. 
But, because of the victory, the user's experience state is not bad. 
His winning motivation during the game promotes a good user experience state. He also wants to play more games.

\textbf{3D MovFlow} We propose a three-dimensional MovFlow (GUT) theoretical model composed of motivation, skills, and challenges (respectively corresponding to the Z, Y, and X axes of the Cartesian coordinate system that meets the right-hand rule).
A good MOBA game has characteristics that engage players to improve their skills, take on more difficult challenges, and have stronger motivations. 
Thus, we believe that in the three-dimensional field, each entity (axis) gives a positive impetus to the player’s GUX state.
These forces cause a perturbation phenomenon to the player's GUX, resulting in the instability of the GUX.
We can judge the player's GUX state by observing the direction of the movement trend generated by the combined forces of the three dimensions in space at a certain moment.
In the law of gravity \cite{newton1987mathematical}, the magnitude of the force is proportional to the masses of the objects and inversely proportional to the square of the distance between them, as illustrated in Equation 1. However, in the GUX Field, the mass of GUX is stable, so we let the numerator of Equation 1 equal the constant $K$, which is determined by the game characteristics, as illustrated in Equation 2. While the constant $K$ of the three-axis should be the same.

\begin{equation}
\centering
\mathrm{F}=\frac{\mathrm{GM} \mathrm{m}}{\mathrm{r}^{2}}
\end{equation}
\begin{equation}
\mathrm{F}=\frac{\mathrm{K}}{\mathrm{r}^{2}}
\end{equation}
, where the $r$ means the distance from the point to the axis.

We included the motivation dimension to the original Flow model and found that when the player's GUX is in a state of balance between Flow and motivation (the player's skill, challenge, and motivation are in balance), the player enjoys the positive effect of positive affect (motivation) on the user experience, or he will enter a state of exhilaration or frustration due to too much or too little positive affect, which interferes with the player's game experience. There is a central axis $A$ in the GUT model, which has the general equation X (Challenge)=Y (Skill)=Z (Motivation). When GUX is located on this central axis, the distance from GUX to the three axes is equal, the non-axial combined external force of the three dimensions on GUX is 0, and GUX is in a non-axial stable state. The GUX on the axis has a transient motion tendency toward more difficult challenges, higher skill, and stronger motivation. This is consistent with the findings of the game user research mentioned above. Thus, the best GUX state, GUT=2 is located on the central axis $A$ in the GUX Field.

According to the Experience-Flow model (Equation 3) proposed in the work \cite{moneta2012measurement}, where the $E_p$ denotes experience and $X_C$ and $Y_S$ are Challenge and Skill, respectively, if the regression coefficients of challenge $\beta_{1}$ and skill $\beta_{2}$ are both positive and of equal value, and $\beta_{3} \le 0$, then the quality of experience is the highest in the Flow channel. In the GUT model, there should be a Tunnel with the central axis $A$ as the axis, which has the characteristics of overlapping with the current Flow model when projected onto a two-dimensional plane as illustrated in Figure 2. It contains all good GUX (GUT=1, 2) in the GUX Field.

\begin{equation}
  E_p=\beta_{0}+\beta_{1} X_C +\beta_{2} Y_S + \beta_{3} \text{|}X_C-Y_S\text{|}  \text{,}
\end{equation}
Therefore, we propose the GUT equation Equation 4 when the player's motivation dimension has not reached its peak as follows:
\begin{equation}
 \left(\beta_{1}(X_{C}+Y_{S})+\beta_{2} Z_{E}+C\right)^{2} +\beta_{3}(X_C+Y_S)^{2} \leq C^{2} \text{,}
\end{equation}
, where the $X_C$, $Y_S$, and $Z_E$ denote Challenge, Skill, and Motivation, respectively. In the ideal case where $\beta_{1} =\beta_{3} > 0$ , and $\beta_{2}<0$ , the player is in the good GUX state. 
According to the prior knowledge in the field of game planning, we consider that the distribution of the GUX value in the Skill-Challenge-Motivation space basically follows the normal distribution. 
Thus, the GUX value $l$ follows a bell-shaped curve on each tangent plane of the Tunnel in the space
and its maximum value is located on the central axis $A$. Let the coordinates of a point $P$ in the GUX Field be ($x_1$,$y_1$,$z_1$) then the distance from the point $P$ to the central axis $A$ is:

\begin{equation}
\begin{split}
    &d=\frac{\left|\left\{-x_{1},-y_{1},-z_{1}\right\} \times\{1,1,1\}\right|}{\sqrt{1^{2}+1^{2}+1^{2}}}\\&=\frac{\sqrt{\left(z_{1}-y_{1}\right)^{2}+\left(x_{1}-z_{1}\right)^{2}+\left(y_{1}-x_{1}\right)^{2}}}{\sqrt{3}} \text{,}
\end{split}
\end{equation}
and the GUX value "l" of point $P$ is :
\begin{equation}
l=\frac{1}{\sqrt{2 \pi}} e^{-\frac{d^{2}}{2}}=\frac{1}{\sqrt{2 \pi}} e^{-\frac{\left(z_{1}-y_{1}\right)^{2}+\left(x_{1}-z_{1}\right)^{2}+\left(y_{1}-x_{2}\right)^{2}}{6}} \text{,}
\end{equation}
, where the value of the good GUX should meet the Equation 7 condition. The $C$ in Equation 7 and 4 is the radius of the tunnel. 
\begin{equation}
l \ge \frac{1}{\sqrt{2 \pi}} e^{-\frac{C^{2}}{2}} \text{,}
\end{equation}

We define the user experience state that is, in the region from the central axis to the surface of the cylinder as the GUT=1 state, and outside of the cylinder's surface will be GUT=0 state.
We use equations 8, 9 and 10 to calculate the values of the motivational dimensions using fuzzy logic.

\begin{equation}
\begin{array}{l}
\mathrm{E}_{\mathrm{p}}^{\mathrm{i}}=1, i=1,2, \cdots, m \\
\mathrm{E}_{\mathrm{N}}^{\mathrm{j}}=1, \mathrm{j}=1,2, \cdots, n \text{,}
\end{array}
\end{equation}
Also, there will be $m$ positive and $n$ negative affects
\begin{equation}
\begin{split}
    &\# \mathrm{PA}=\sum_{i=1}^{\mathrm{m}} \mathrm{E}_{\mathrm{P}}^{\mathrm{i}} \\
&\# \mathrm{NA}=\sum_{i=1}^{\mathrm{n}} \mathrm{E}_{\mathrm{N}}^{\mathrm{i}} \text{,}
\end{split}
\end{equation}

\begin{equation}
\delta= \# \mathrm{PA}-\# \mathrm{NA}-\mathrm{D} \text{,}
\end{equation}
, where $\delta$ is the value of the motivation dimension. The value of $D$ is determined by the fuzzy decision part in section 2.3, which is $D = 2$.

\subsection{GUXAS Multi-modal Data Acquisition System}

To reduce the psychological pressure user's go through when collecting their physiological data, we built a wearable multi-modal physiological data collection system. This system collects the users' EEG, ECG, and GSR physiological signals, audio (voice emotion) and video (facial emotion) during the game play in real-time, as illustrated in Figure 5. 

\begin{figure*}
  \centering
  \includegraphics[width = \textwidth]{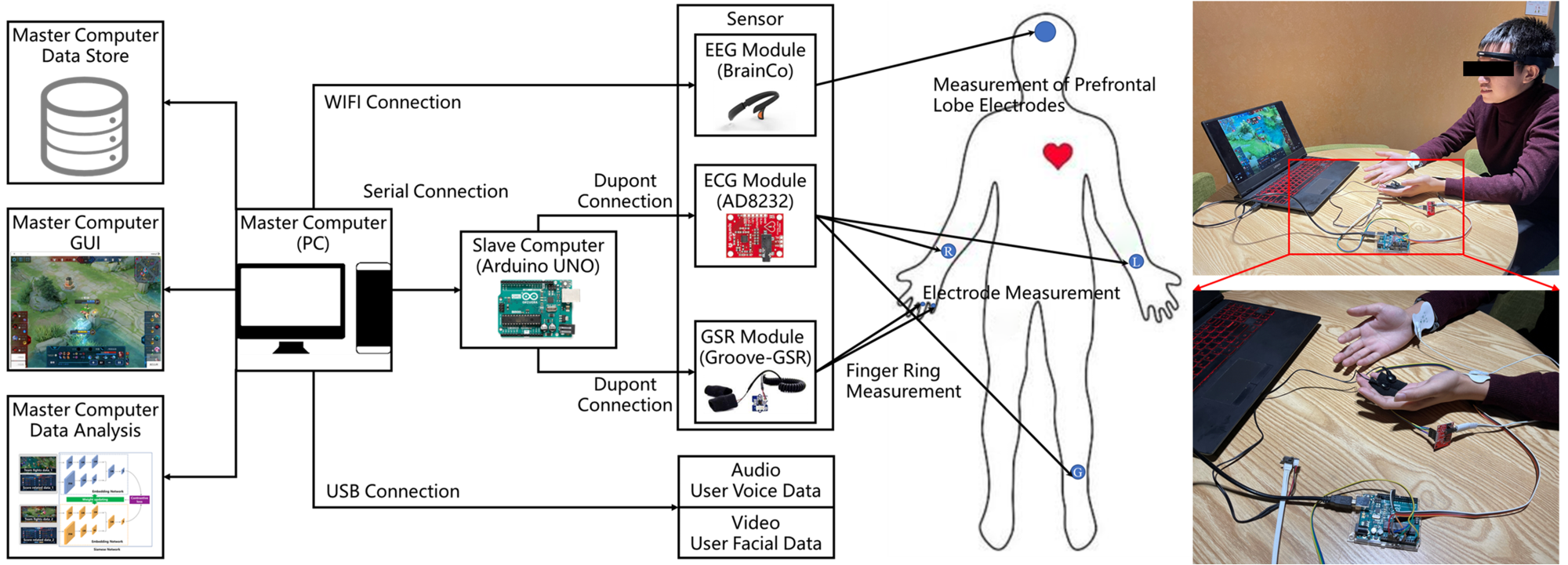}
  \caption{
The left side is system link graph, while the right side is the person undertaking the Q\&A experiment wearing our multi-modal system. The system consists of Groove-GSR, AD8232 and BrainCo, which collect the EEG, ECG and GSR signal of the players.
}
\end{figure*}

We design and implement the ECG module assembling with lightweight makers based on Arduino and AD8232, \\which is an integrated signal conditioning module. The device is designed to extract, amplify and filter weak bio-electric signals in the presence of noise caused by movement or remote electrode placement.  
This design enables an ultra-low power Analog-to-Digital Converter (ADC) or embedded microcontroller to easily collect the output signal. AD8232 amplifies the ECG voltage from 0 to 5V, and Arduino reads the value through its analog input pin. The Arduino controller has multiple 10 digit A / D channels. This means that Arduino can map the voltage input signal of 0-5V to the value of 0-1023. The data is transmitted back to the computer through the serial port. 
GSR module consists of Groove-GSR and Arduino. GSR is a measurement method between skin conductance. Affects will stimulate the sympathetic nervous system, causing the sweat glands to secrete more sweat. Groove-GSR can find affect waves by simply connecting two electrodes, two fingers of one hand. 
Arduino is open-source hardware often used in lightweight maker designs.

According to the Shannon–Nyquist sampling theorem \cite{RAVANSHAD2020223}
in the process of analog/digital signal conversion, the sampling frequency should be greater than or equal to twice the highest frequency of the original signal, so that the digital signal after sampling can completely retain the characteristic information in the original signal. The average number of human heartbeats is generally about 90 beats per minute, 1.5HZ. In order to accurately collect the signal characteristics of the player heartbeat, we set the signal collection frequency of the sympathetic nerve to 182HZ.

We use the BrainCo headband called Focus 1 to collect the EEG signal. This is a lightweight EEG device with 3 hydro-gel electrodes. The headband uses WiFi to collect data at a frequency of 160.6 Hz. Combined with the Focus SDK, it can extract multi-dimensional EEG related signals such as alpha, beta, gamma, attention and meditation. The Brainco headband use FPZ and TP9 positions in the standard 10-20 international system \cite{homan1987cerebral}.

\subsection{Deep DTW algorithm for dynamic classification of user experience states}

We use DTW\cite{DBLP:journals/corr/abs-1003-4083} matching algorithm with a sliding window
\cite{fritz2014using} for deep learning optimal correlation sequence analysis method to calculate the player's GUX state. We use the physiological data collected from the induce experiment and multi-modal acquisition experiment, to determine the user's experience during the game play as illustrated in Figure 4.

The raw physiological data needs to be de-noised and normalized first.
We extract the following 6-dimensional features from the de-noised physiological data. The features extracted from each physiological data are presented in Table 1. These features include the mean ($\mu$), median, variance ($\sigma$), maximum, minimum, and range ($\delta$) of EEG, ECG, and GSR signals \cite{2004Emotion}. Furthermore, by detecting the distance change between the RR waves (RR interval) in the ECG signal, the HRV signal \cite{LANFRANCHI2017142} can be obtained \cite{2014Matlab}. We also extract the mean value of HRV signal and other common statistical characteristics of HRV such as SDNN (overall standard deviation), SDANN (standard deviation of the mean), RMSSD (square root of the mean square of difference)\cite{2011Discrimination,ye2020flow}, as part of the feature matrix. 

The EEG signal has 6 dimensions including alpha, delta, gamma, high-beta, low-beta, and theta,  which can represent certain activity characteristics of the human brain at the moment. We can also obtain the Attention and Mediation signal from BrainCo headband \cite{kosmyna2019attentivu}. 

\begin{table*}
\centering
  \caption{Features extracted from physiological data. The ECG data is collected from the participants' arms and lower legs using gel electrode pads. The EEG data is Collected from FPZ and TP9 positions in the standard 10-20 international system. The EEG data includes alpha, delta, gamma, high-beta, low-beta, and theta. It also includes the calculated signal Attention and Meditation. The GSR data is collected from the participant's ring finger and little finger to avoid disturbing the normal game play. The HRV data is calculated by ECG data \cite{2014Matlab}}
  \label{tab:commands}
  \begin{tabular}{ccc}
    \toprule
Physiological data &  Feature name & Extracted feature \\
    \midrule
\multirow{6}{*}{ECG} & $\mu_{ECG}$ & Mean of ECG  \\
                     & median$_{ECG}$ & Median of ECG \\
                     & $\sigma_{ECG}$ & Variance of ECG \\ 
                     & max$_{ECG}$ & Maximum of ECG \\
                     & min$_{ECG}$ & Minimum of ECG\\
                     & $\delta_{ECG}$ & Range of ECG \\ \hline
\multirow{6}{*}{EEG} & $\mu_{EEG}$ & Mean of EEG \\
                     & median$_{EEG}$ & Median of EEG  \\
                     & $\sigma_{EEG}$ & Variance of EEG \\ 
                     & max$_{EEG}$ & Maximum of EEG \\
                     & min$_{EEG}$ & Minimum of EEG\\
                     & $\delta_{EEG}$ & Range of EEG \\       \hline              
\multirow{6}{*}{GSR} & $\mu_{GSR}$ & Mean of GSR  \\
                     & median$_{GSR}$ & Median of GSR  \\
                     & $\sigma_{GSR}$ & Variance of GSR \\ 
                     & max$_{GSR}$ & Maximum of GSR \\
                     & min$_{GSR}$ & Minimum of GSR\\
                     & $\delta_{GSR}$ & Range of GSR \\ \hline
\multirow{6}{*}{HRV} & max$_{HRV}$ & Maximum of HRV \\
                     & min$_{HRV}$ & Minimum of HRV \\
                     & SDNN$_{HRV}$ & \tabincell{l}{Standard deviations of all N-N intervals} \\ 
                     & SDANN$_{HRV}$ & \tabincell{l}{ Standard deviation of the means of several 
                                          N-N intervals \\ segmented by a specific time session} \\
                     & RMSSD$_{HRV}$ & \tabincell{l}{Root-mean-square of the difference between adjacent
N-N intervals}\\
                     & $\mu_{HRV}$ & Mean of HRV \\                     
    \bottomrule
  \end{tabular}
\end{table*}


We get three kinds of feature matrix, which are Video feature matrix, Whole feature matrix, and Sliding feature matrix. The physiological data collected from the induced experiment is used to generate the Video feature matrix, while the one collected from the multi-modal acquisition experiment will be used to generate the Whole feature matrix and Sliding feature matrix. The difference between the Whole and Sliding feature matrix is that, the Sliding feature matrix is extracted from a sliding window with a step size of 1 second and a window size of 10 seconds while the data source of the Whole feature matrix is the physiological data collected during the whole game. Among the three matrices, the Whole feature matrix and the Video feature matrix have the Flow label and the affect label, respectively. 
We use the DTW algorithm to calculate the distance between the Sliding feature matrix with the Whole feature matrix, and the Video feature matrix, to determine the GUX label of the user's physiological data in real-time.

The experiment demonstrated that when players have similar physiological data characteristics, they will also have the same Flow experience. Thus, at one moment, if the characteristics of the Sliding feature matrix match with the Whole feature matrix, we consider that the player has the same Flow state as the whole game state and vice versa, at that moment. We use the DTW algorithm to get the corresponding distance between the Video Feature Matrix and the Sliding Feature Matrix. Through a two-layer FC and logistic regression, we obtained the Positive Affects (PA) and Negative Affects (NA) list of the game process.
The Flow list is obtained with a similar method, where the Video feature matrix is replaced with the Whole feature matrix.

Having got the Positive Affects (PA), Negative Affects (NA), and Flow States list, we calculate the Pearson correlation (equation 11) between the PA, NA, and Flow list to get the PA correlation coefficient $ X_1 $ and the NA correlation coefficient $ X_2 $. According to the classic Flow theory, there is a positive significant relationship between Flow and PA \cite{mundell2000role}. At the same time, there is no correlation between Flow and NA \cite{article}. We design our loss function: $ X_1-X_2 $ $\leq$ C ($ \lim_{X_1 \to 1}$) according to Pearson's correlation coefficient. When the correlation coefficient is less than or equal to 0.4, it is weakly correlated or not correlated. So we assume that $C$ is 1.4 in this paper. After estimating the loss function, we use the gradient descent back-propagation method to continuously iterate to get the optimal network parameters. We define GUT=0, 1, 2 as three different GUX states representing the best GUX state, good GUX state, and average GUX state in our model. 
We use equations 8, 9 and 10 to calculate the values of the motivation dimensions and use them as input to the fuzzy logic. Based on the fuzzy logic flowchart shown in Figure 6, we quantified the player's game user experience state as follows.
If the player is not in the state of Flow, he will be in the GUT=0 state in our model. When the player is in the Flow state, with motivation value $\delta$ of player's equal 0 and the correlation coefficient $X_1 \ge 0.6$ then, the GUX state is GUT=2, otherwise the GUX state is GUT=1.  

\begin{figure*}
  \centering
  \includegraphics[width = \textwidth]{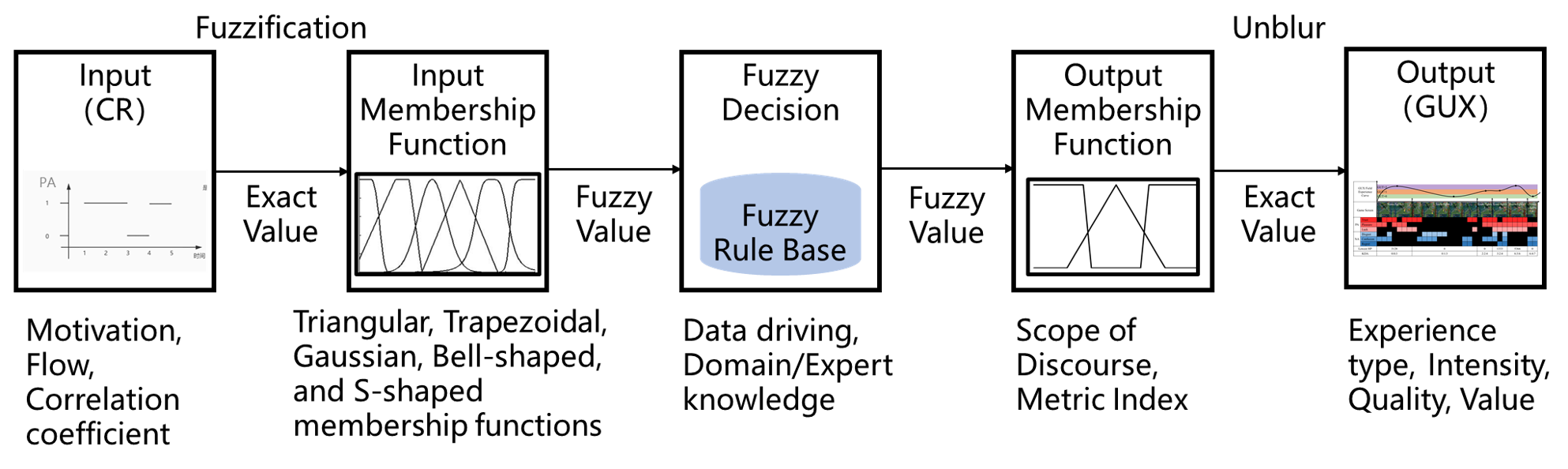}
  \caption{
Where the CR means the Classification Results. It consists of the four components of the fuzzy interface, the fuzzy decision module, the fuzzy solution module, and the fuzzy rule base in the fuzzy inference system.
The model consists of four components: input, input membership function, output membership function and output rule base. We will combine expert knowledge in the domain with a data-driven approach way to unblur the fuzzified classification results, and ultimately output the GUX state of the users.
 }
\end{figure*}

\subsection{Online user experience state prediction algorithm based on Siamese Network}

\textbf{Game Process Data}
We divide the game process data into three categories: Team fights data, Score related data and in-game hero coordinates. The Team fights data has 11 dimensions, which is the game process data during the team fights, including the player's blood volume, damage and other information during the team fights. The Score related data has 33 dimensions, which mainly includes information about the player's match time, hero level, coins, kills, during the whole match. 

Due to the limitations in exporting game process data, the data granularity is too large to analyze in real-time. Therefore, we mainly focus on the main factors that affect the results of matches in MOBA games, which include killing important map mechanisms (such as master, tyrant) and team-fights. However, according to the game process data collected, the important mechanisms like purchasing gears are discrete data that cannot be used as the main features for model training. So, we analyzed the player's GUX state during the outbreak of the team-fights. 

According to the GUX state (GUT=0, 1, 2)  obtained by the Deep DTW matching algorithm,  we found that players have different GUX intensities and lengths when a battle breaks out. Due to the small samples of the dataset available, we choose the few-shot learning as our main method.
The few-shot learning mainly includes transfer learning, meta-learning, and metric learning.
Deep learning is barely used in this field, hence, no pre-trained model that suits transfer learning.
At the same time, the lack of sub-tasks and data cause Meta-learning is also not suitable.
Therefore, we chose metric learning based on the Siamese network \cite{chopra2005learning}, which has the characteristics of not requiring pre-trained models, fewer data requirements, and capable of better classification performance.

\begin{equation}
r=\frac{N \sum x_{i} y_{i}-\sum x_{i} \sum y_{i}}{\sqrt{N \sum x_{i}^{2}-\left(\sum x_{i}\right)^{2}} \sqrt{N \sum y_{i}^{2}-\left(\sum y_{i}\right)^{2}}}  \text{,}
\end{equation}

\begin{algorithm}[H]
  \caption{Deep DTW Affect labeling}
  \label{alg:1}
\begin{algorithmic}
  \STATE {\bfseries Input:} Physiological Data of Video Induce $PV$, Physiological Data of Game Process $PG$
    \STATE {\bfseries Output:} Positive and Negative Affects List $PNL$,Flow List $FL$
  \STATE Get de-noised $PV$ and de-noised $PG$
  \STATE Extract features from the de-noised $PV$ and de-noised $PG$ based on sliding window
  \STATE Get DTW matching result sequences from Video Feature Matrix and Sliding Feature Matrices $S_{vsms}$, Whole Feature Matrix and Sliding Feature Matrices $S_{wsms}$
  \STATE Initialize parameters $W$ randomly
  \FOR{$i=1$ {\bfseries to} $MaxSteps$}
  \STATE Feed $S_{vsms}$ and $S_{wsms}$ to Neural Network
  \STATE Calculate the loss function via formula 
  \STATE Back propagation and update parameters $W$ 
  \ENDFOR
  \STATE Feed $S_{vsms}$ and $S_{wsms}$ to Neural Network to get $PNL$ and $FL$
  \STATE return $PNL$ and $FL$ 
\end{algorithmic}
\end{algorithm}

During the team-fights, we found that most players are in the state of GUT=1 and GUT=2 (This is consistent with the real feelings of players during the game-play which we got through interviews), which causes data imbalance problem. To reduce the individual differences which influence the results, we propose a Personality Embedding Network. It is based on the game score including the number of kills, money, and other 30 dimensions data collected at the end of the game and used as the input for the personality coding. Through these score-related characteristics, the personal game tendency can be obtained, and the problem of poor training results caused by individual differences in players will reduce. To sum up, we propose a set of metric learning and few-shot learning methods based on Siamese Network and Personality Embedding Network to establish a mod\\-el between game data and GUX state, which is illustrated in Figure 7.

\begin{figure*}
  \centering
 \includegraphics[width = \textwidth]{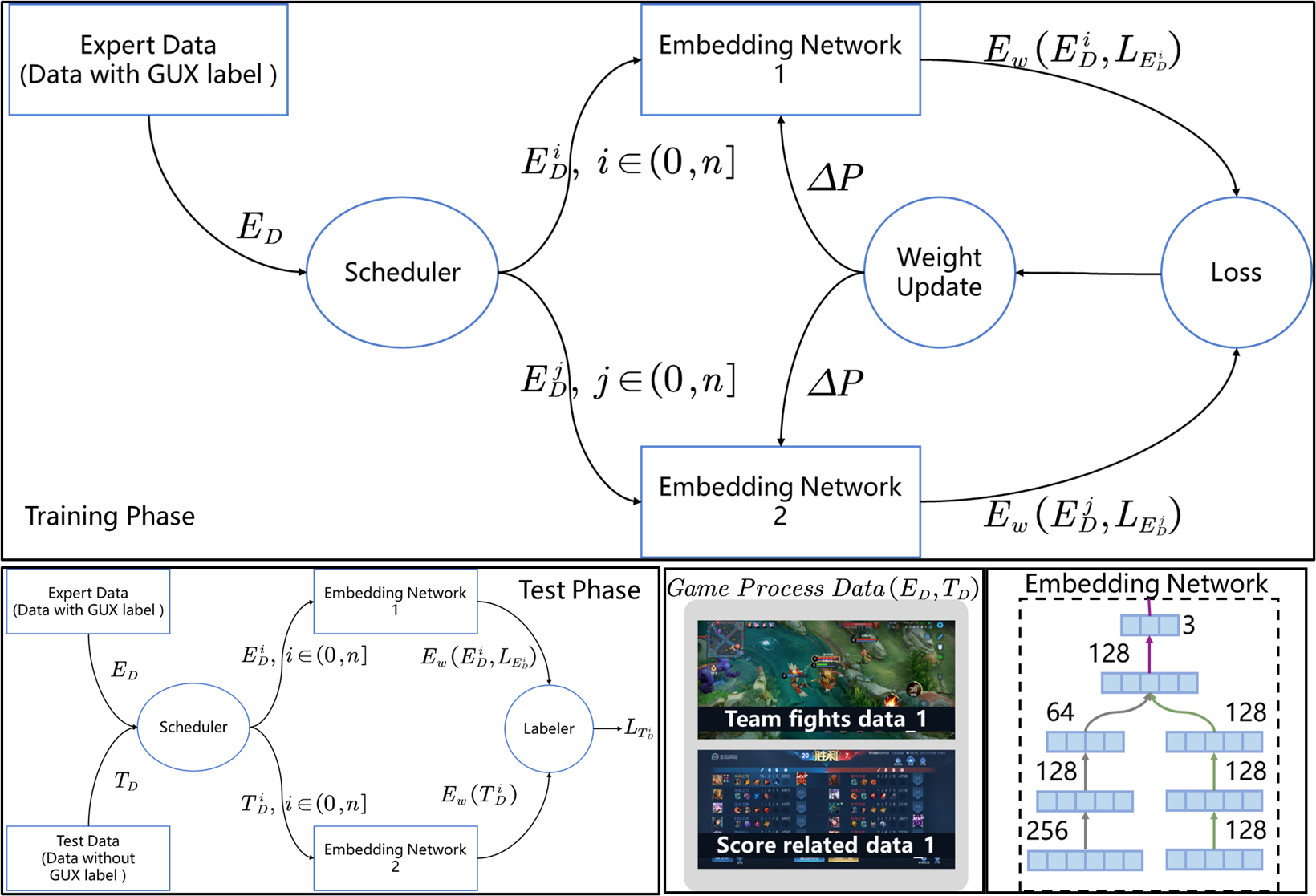}
  \caption{
GUX state prediction network, which is a Siamese network composed of two identical Embedding networks. We use the score related data to judge the players’ skill, challenge, and mentality throughout the game. To a certain extent, we infer the personal tendency of the users during the game (aggressive, conservative, whether they like team-fights or not) as the personal characteristics of each player. During the training phase, we will input the game process data with GUX label into the scheduler. It will randomly match a pair of data, which are $E_{D}^{i}$ and $E_{D}^{j}$. The two embedding networks have the same network structure and will share parameters. During the test phase, we will input input the test data and expert data into the scheduler, and we can obtain the label of the test data.}
\end{figure*}

Firstly, we tried to solve the classification problem with a simple embedding network.
We pass the input game data features (11-dimensional Team fights data, 33-dimensional Score related data) through a series of fully connected networks. The output data (Output, 3-dimensional) is entered into the Softmax function to get the probability of p(GUT=0), p(GUT=1), p(GUT=2). The argmax of the probability obtained above will be the predicted GUX state.  Because there are fewer samples of GUT=0 and GUT=2 with labels, we set a high loss when the ground truth is GUT=0 or GUT=2. The loss function of embedding network is weighted cross-entropy ($weight_l=[2, 1, 2]$), $y^{\text{gt}}$ is ground truth and the $p_{c}^{'}$ is the probability of predicting class C, as shown in equation 12. In addition, we use the oversampling technique to mitigate the data imbalance problem.

\begin{equation}
L\left(x, y^{\text {gt }} \text{,} \theta\right)=-\sum_{c=0}^{2} \text { weight }_{l} * y_{c}^{\text {gt }} * \log \left(p_{c}^{'}\right) \text{,}
\end{equation}

Although, we use several techniques to balance the imbalance data, it is still easy to over-fit on the training set, during the multi-classification process. Therefore, we choose metric learning. Two output results from two different input data are obtained from the Embedding network with shared parameters and then, we will measure the distance between the two Outputs.The input of the embedding network we use in Siamese network is game data features with GUT label (11-dimensional Team fights data, 33-dimensional Score related data). The distance between the outputs of two similar data samples is low and vice versa when data samples are parallelly passed to the Embedding network. We use contractive loss function to evaluate the outputs of the Embedding network as shown in equation 13.

\begin{equation}
\begin{split}
    &L\left(X_{1}, X_{2}, Y, W\right)= \\ & \frac{1}{2}\left(Y * D_{w}^{2}+(1-Y) * \left(\max \left(0, \operatorname{margin}-D_{w}\right)\right)^{2}\right) \text{,} \\
&D_{w}=|| E_w(X_{1})-E_w(X_{2})||_{2}  \text{,}
\end{split}
\end{equation}

Where $E_w$ is the Embedding network, $W$ is the network weight; Y is the paired label. If the pair of samples $X_1$ and $X_2$ belong to the same class, Y=1, on the contrary, Y=0. $D_w$ is the Euclidean distance between $E_w(X_1)$ and $E_w(X_2)$. 
When Y=1, adjust the parameters to minimize the distance between $E_w(X_1)$ and $E_w(X_2)$. When Y=0, then the distance between $E_w(X_1)$ and $E_w(X_2)$ is greater than the margin. Meaning the distance between the two samples is relatively wide and no need to optimize it. If the distance between $E_w(X_1)$ and $E_w(X_2)$ is less than the margin, then we need to increase the distance to the margin.

We randomly assign all the game process data according to 3:7, with 70 $\%$ being the training set and 30 $\%$ being the test set. When we make predictions, we need to use the training set to obtain the mean of each class of output as $baselines_{0, 1, 2}$. Pre-processing the test set data gives us the $Output_{test}$. Calculate the distance (similarity) between $Output_{test}$ and $baselines_{0, 1, 2}$. The GUX label of $Output_{test}$ will be as same as the label of $baselines_{0, 1, 2}$, when the distance between $Output_{test}$ and $baselines_{0, 1, 2}$ is the minimum.


\section{Experiment Setup}

In this section, we introduce two experiments we designed in detail, including the induce experiment (short as Q$\&$A Interviews ) and multi-modal acquisition experiment (short as MOBA Game Playing). The two phases in total will take about 1 hour to complete. We get the affect label of the physiological data feature matrix in the phase of questionnaire surveying, and the multi-modal data which includes the physiological and game process data in the phase of Game playing. In both of the two phase, we will collect the physiological data from the participants.

\begin{figure*}
  \centering
\includegraphics[width=\textwidth]{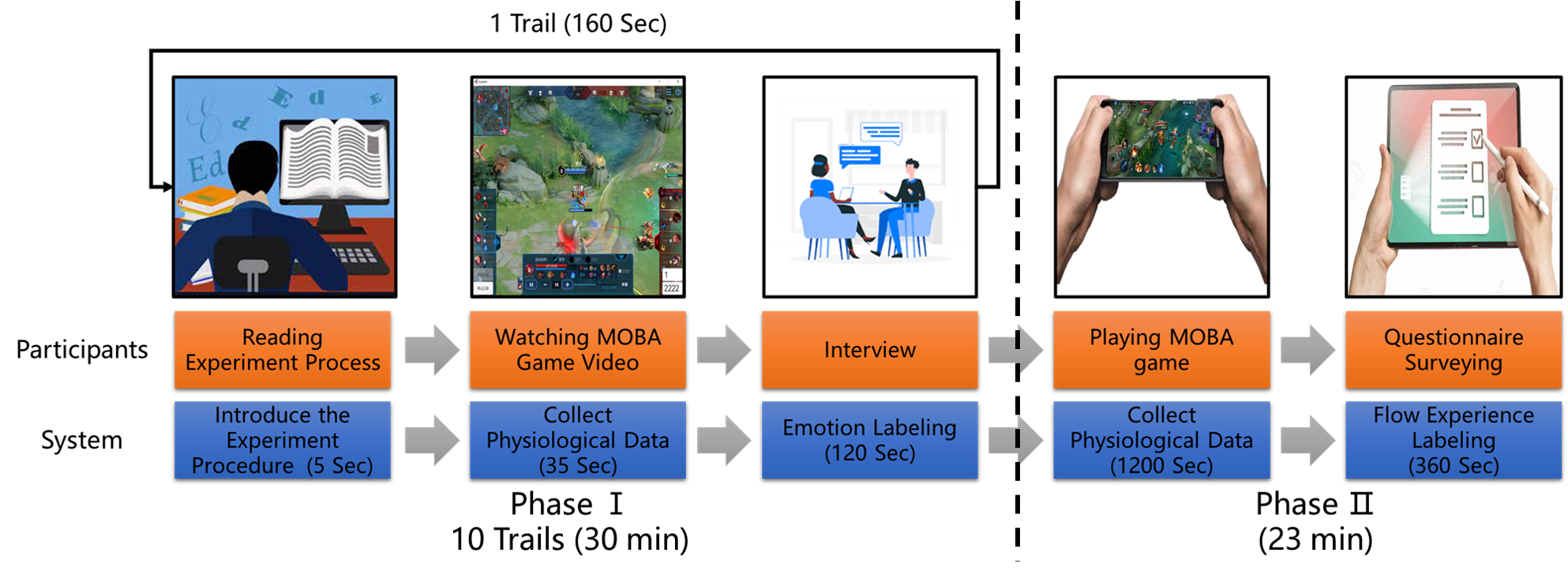}
  \caption{Experiment Pipeline. Where Q$\&$A Interviews (induce experiment) is Phase \uppercase\expandafter{\romannumeral1} and MOBA Game Playing (multi-modal acquisition experiment) is Phase \uppercase\expandafter{\romannumeral2}  }
\end{figure*}

\subsection{Phase of Question and Answer Interviews} 


We chose MOBA game videos as emotion-evoking materials. 
After pre-processing, we obtained a total of 76 videos that could induce affects in the participants, and they could induce seven affects, such as pleasure, surprise, trust, lucky, confusion, regret and disgust. 
In this stage, participants are required to watch 10 emotion-evoking videos and label them with affect labels. 
The labeled labels are assigned to the physiological data collected during watching the videos. The Flow of the affect labelling experiment consisted of three parts, as shown in Figure 8 Phase \uppercase\expandafter{\romannumeral1}.
At the beginning of the experiment, the player is reminded that the video is about to start and will show the identification card with the video number for 5 seconds. 
The emotion-evoking video of 35 seconds in length will then be played, and while it is playing, our physiological data collection system will capture physiological data from the participant watching process. 
Next, participants will be asked to conduct a semi-structured interview to label the video with affect label.
Interviews will be conducted with participant in conjunction with questions from the SAM \cite{bradley1994measuring} and PANAS questionnaire \cite{watson1988development} to confirm their emotional state during the experiment.

\begin{algorithm}[H]
  \caption{GUT State Prediction}
  \label{alg:2}
\begin{algorithmic}
  \STATE {\bfseries Input:} Game Play Data $X$, Flow Label $Y$
    \STATE {\bfseries Output:} Flow Prediction $Y^{pred}$
  \STATE Initialize parameters $W$ randomly.
  \STATE Extract team-fights and score-related data
  \STATE Split dataset to training set and test set
  \FOR{$i=1$ {\bfseries to} $MaxSteps$}
  \STATE Make pairs data from training set for siamese network
  \STATE Feed pairs data to Embedding Network
  \STATE Calculate the loss function via formula 
  \STATE Back propagation and update parameters $W$
  \ENDFOR
  \STATE Get the mean of output of class 0, 1, 2 from training set as $baselines$
  \STATE Feed test data to Embedding Network to get $Output_{test}$
  \STATE Calculate the distances between the $Output_{test}$ and $baselines$
  \STATE Choose the class to which the minimum distance belongs as Flow prediction $Y^{pred}$
  \STATE return Flow Prediction $Y^{pred}$
\end{algorithmic}
\end{algorithm}

\subsection{Phase of MOBA Game playing.}
 The phase of game-play testing includes two stages, which are Playing MOBA Game Stage and Questionnaire Surveying Stage, as illustrated in Figure 8 Phase \uppercase\expandafter{\romannumeral2}. Before the first phase, it is the participant preparation, in which the participant needs to relax and concentrate on the coming game for 30 seconds. In the phase of game-playing, the physiological multi-modal data of the participant is collected. Then, the participant will be asked to take two sets of questionnaires on labeling their Flow state of the entire game procedure. The two questionnaires are the Flow State Scale (FSS) \cite{jackson1996development} and MOBA GameFlow State Scale (HOKGFSS). The FSS questionnaire uses the standard 5-Likert scale to judge the player’s experience status through 36 questions. The HOKGFSS is a set of 35-question questionnaires we designed based on the GameFlow theory \cite{sweetser2005gameflow}, which is mainly used to judge players’ subjective experience of the MOBA Game. During the experiment, the participants did not apply any conductive gel or alcohol to make them as comfortable as they play games at home.
 
 \subsection{Participants}
 Fifty-three participants, including 3 female students and 50 male students, between 19 and 27 years old (M=22.0, SD=1.9) completed the study. 80$\%$ of the participants are experienced MOBA gamers and half of them play MOBA  games at least 10 times a month. All the participants took part in the two experiments and were physiologically monitored. All participants signed the subject consent form, giving their consent for us to collect their physiological data and game process data for data analysis.
 
\section{Results and Analysis}
In this section, we discuss the experimental results from three aspects, which are based on questionnaires, physiological, and game process data.

\subsection{Questionnaire Results and Analysis} 
Although we use the MOBA game video as the emotion-induce data set, which lacks artistic rendering and can only induce players to have relatively weak affects. We still induced 7 kinds of affect in the participants which always appear during the MOBA game-play. These affects include pleasure, trust, surprise, luck, disgust, confusion, and regret. Most of the time, we only induce 6 kinds of affect in the participant, with 3 PAs and 3 NAs. We visualize the results of Q$\&$A Interviews as illustrated in Figure 9 and Figure 10. Our emotion elicitation material is more likely to induce PAs in participants, especially sentiment of trust. We can find the same situation from the questionnaire result of Multi-modal acquisition experiment. Most of the participants we invite can easily generate good GUX state, due to they are the active players of the MOBA game we choose. They have a higher motivation.  

\begin{figure*}
  \centering
 \includegraphics[width = 1\textwidth]{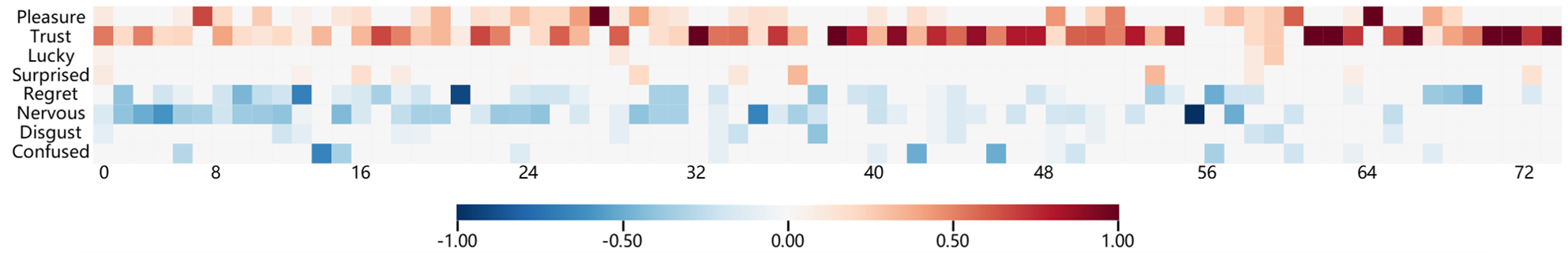}
  \caption{
Results of the heat-map of the induce experiment, with the number of the video on the x-axis, and the sentiment label on the y-axis. Each video was viewed by the same number of participants.
}
\end{figure*}

\begin{figure}
  \centering
 \includegraphics[width = 0.5 \columnwidth]{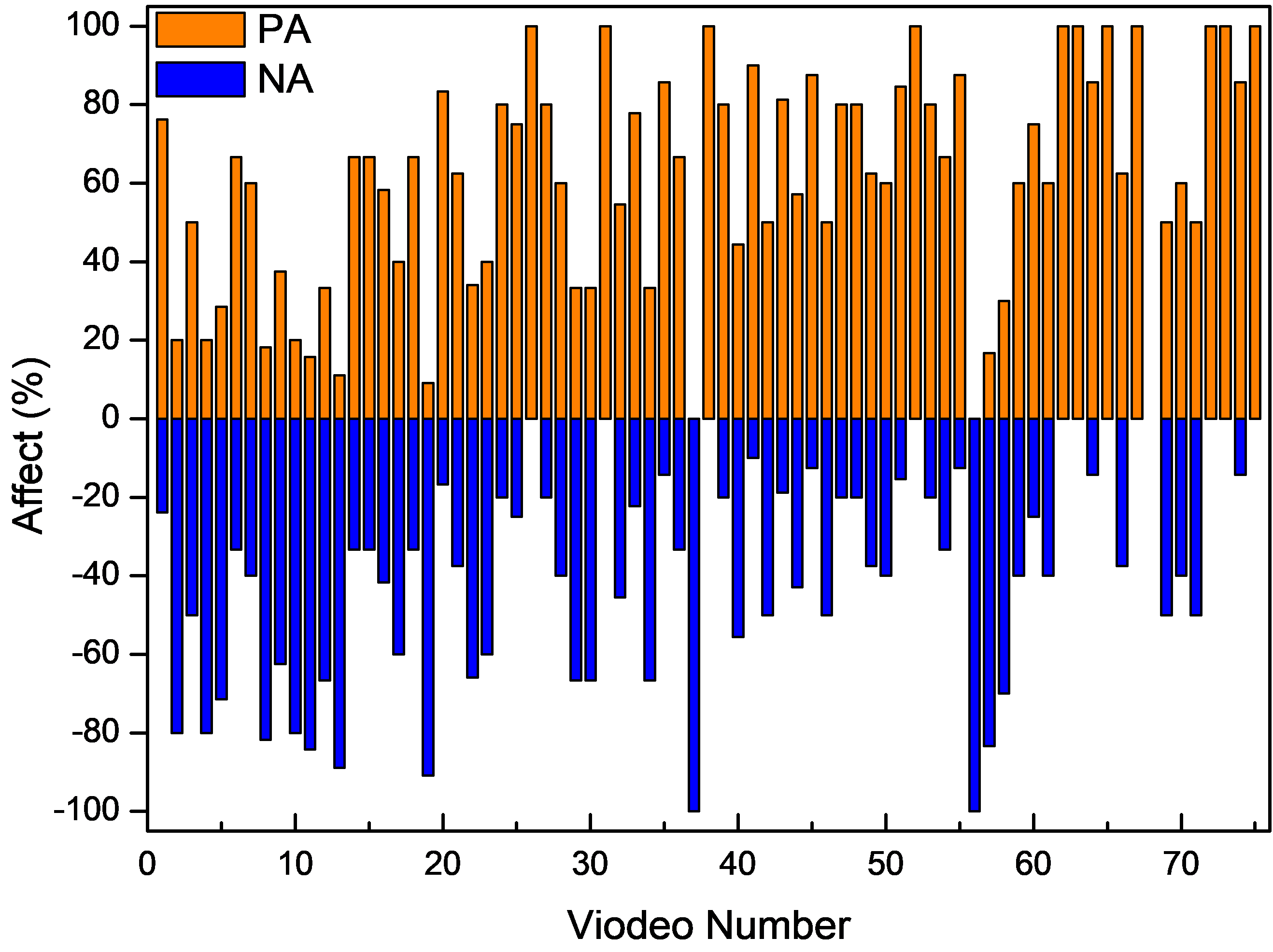}
  \caption{
 Induce experiment affects stacking histogram. Where orange is Positive Affect (PA) and blue is Negative Affect (NA).
}
\end{figure}

We analyze the reliability and validity of the questionnaire to judge whether the questionnaire we used is reliable and reasonable \cite{louangrath2018validity}. We use Cronbach reliability analysis and KMO, Bartlett validity test in SPSS to verify the questionnaires. The results are shown in Table 2.
From the obtained results, it is easy to know that both of our questionnaires are reliable with reliability greater than 0.8. 

\begin{table}
  \caption{Reliability and Validity Test Results of The Questionnaires}
  \label{tab:commands}
  \begin{tabular}{ccl}
    \toprule
Questionnaire name &  Reliability &  Validity\\
    \midrule
FSS &  0.900   & 0.691    \\
HOKGFSS & 0.924 & 0.691  \\
    \bottomrule
  \end{tabular}
\end{table}

Secondly, we try to average the results of the FSS questionnaire because it is using the standard 5-Likert scale \cite{allen2007likert}. If the average of the results is higher than 3, then the participants are in the Flow state ( agree with the topic point of the questionnaire). The average value of the participants' FSS questionnaire results is mostly between 3.2-3.7. After removing outliers, results showed that when participants play the MOBA game, they are mostly in the state of Flow. This situation may be related to the participants in our experiment and the MOBA game itself. Because the result of the HOKGFSS questionnaire shows that all the participants think this game is relatively enough to make them easily stay in Flow state.

\subsection{Physiological Data Results and Analysis} 
We conducted a total of 55 multi-modal acquisition experiments, due to the instability of the WIFI signal and the interruption of the code during the test, only 43 sets of physiological data were collected. Since the physiological data of each participant has individual differences, we trained 43 networks with different parameters to predict the participants' GUX state. We conduct an interview with the participants after MOBA game playing to verify our prediction results from our Siamese network. All participants who participated in the interview experiment agreed with the results we predicted using Deep DTW algorithm.

In these 43 rounds of data, PA are positively correlated with the GUT=2 and GUT=1 states of the GUT model. It suits the Flow theory proposed by Mundell in 2000 \cite{mundell2000role}, which proves that our GUT model has the same properties as Flow.

\subsection{Game Process Data Results and Analysis} 
We use accuracy (ACC), recall, precision, and F1-score as main analysis indicators for the prediction results based on game process data. The normalized confusion matrix of our network and baselines are illustrated in Figure 11, Figure 12. Since we are the first team using MOBA game process data to predict the player's game user experience and few teams using deep learning for emotion estimated, we mainly compared our final results with other ordinary Multi-classification networks, as shown in Table 3. Indicators such as recall are the arithmetic mean of each category, except for accuracy. FC and PE represent Fully Connected and Personality Embedding network. Our GUXAS tool has an additional input, Score Related Data, which comes from the personality embedding network, compared to the normal Siamese network.

\begin{figure}[h]
  \centering
\includegraphics[width= 0.45\columnwidth]{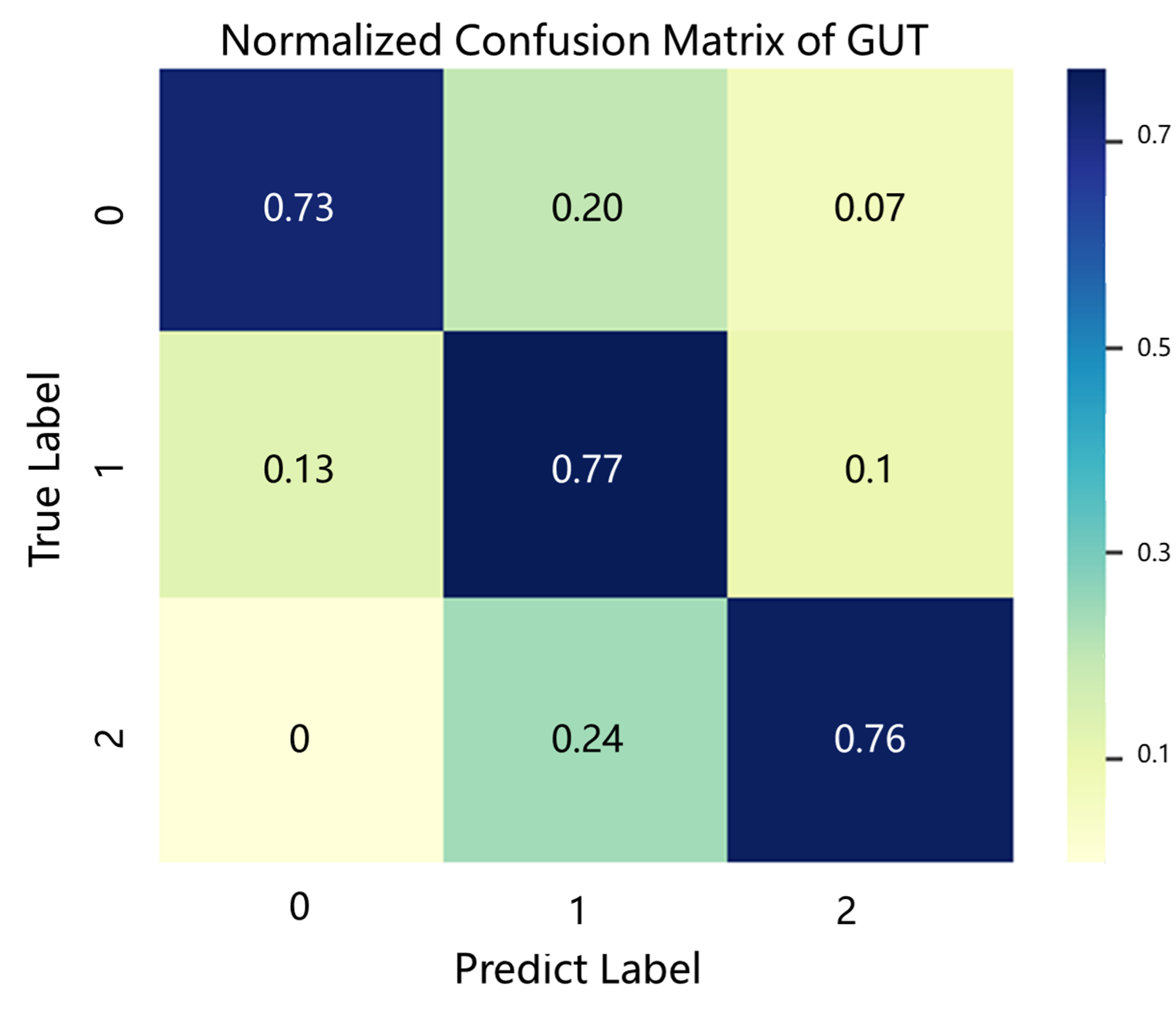}
  \caption{
GUXAS Tool normalized confusion matrix }
\end{figure}

\begin{figure*}[h]
  \centering
\includegraphics[width=\textwidth]{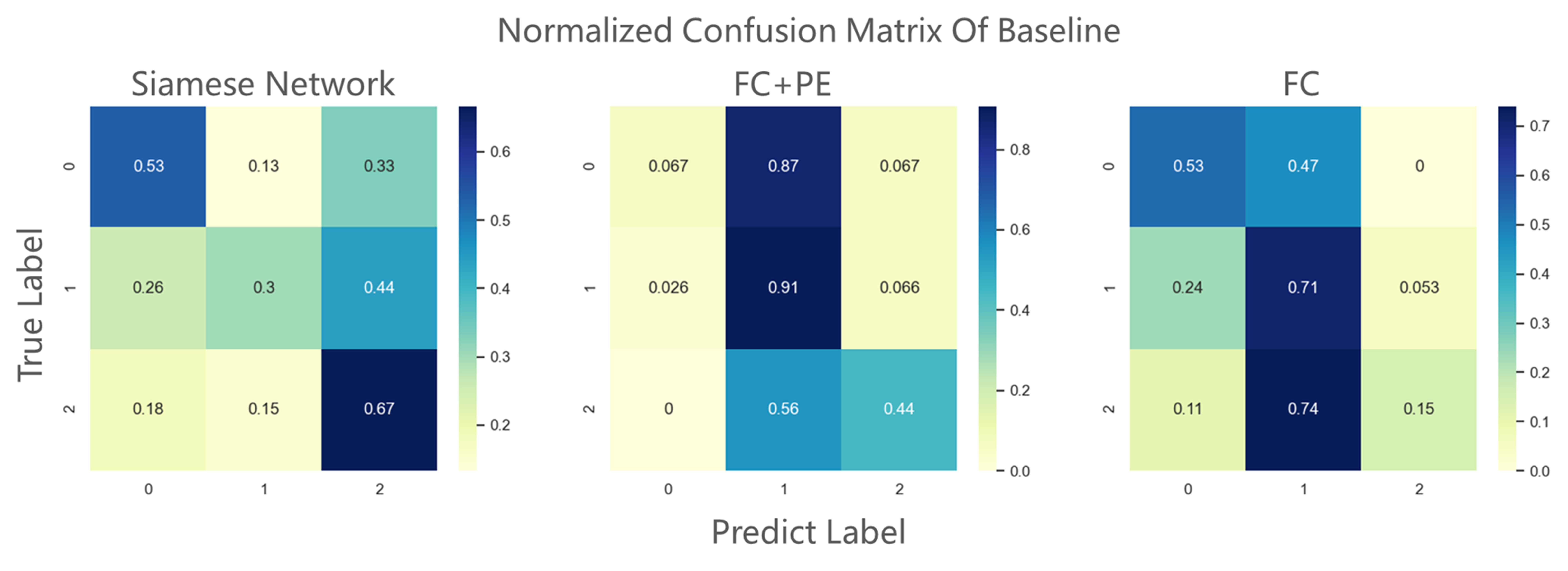}
  \caption{
Baseline normalized confusion matrix }
\end{figure*}

\begin{figure*}
    \centering
    \includegraphics[width=0.46\columnwidth]{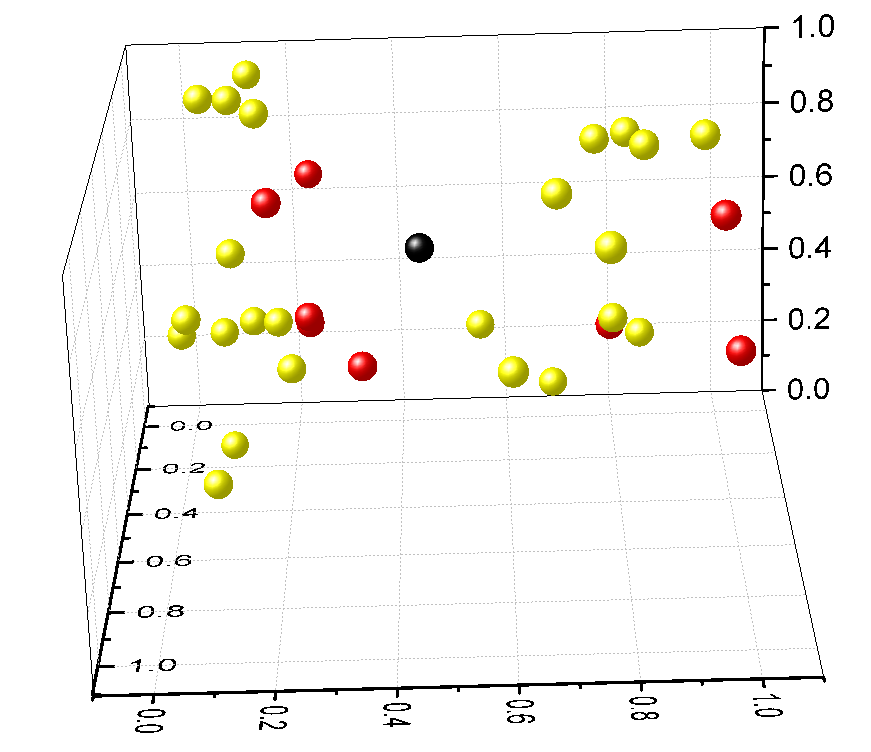}{(a)}
    \includegraphics[width=0.46\columnwidth]{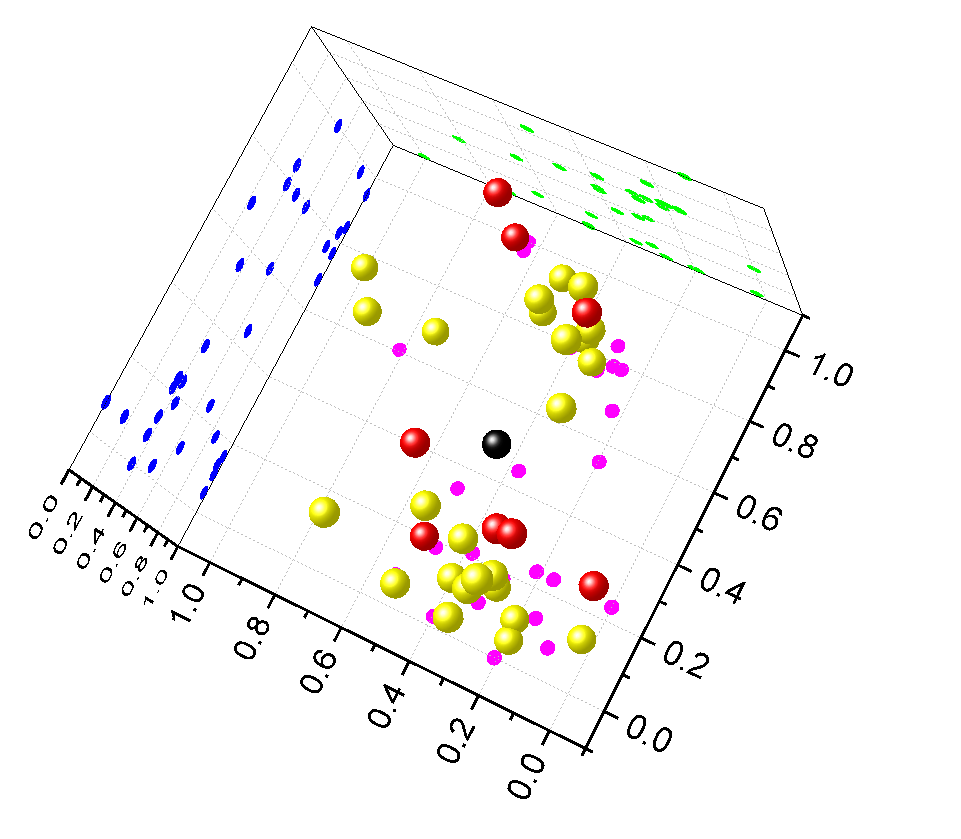}{(b)}
    \includegraphics[width=0.46\columnwidth]{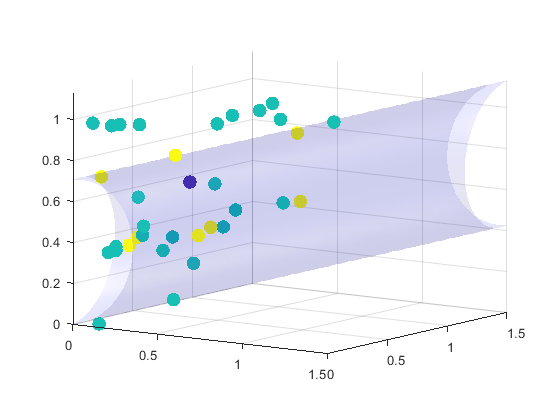}{(c)}
    \includegraphics[width=0.46\columnwidth]{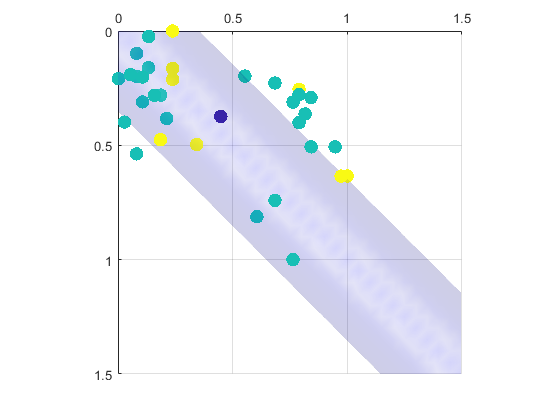}{(d)}
    \includegraphics[width=0.46\columnwidth]{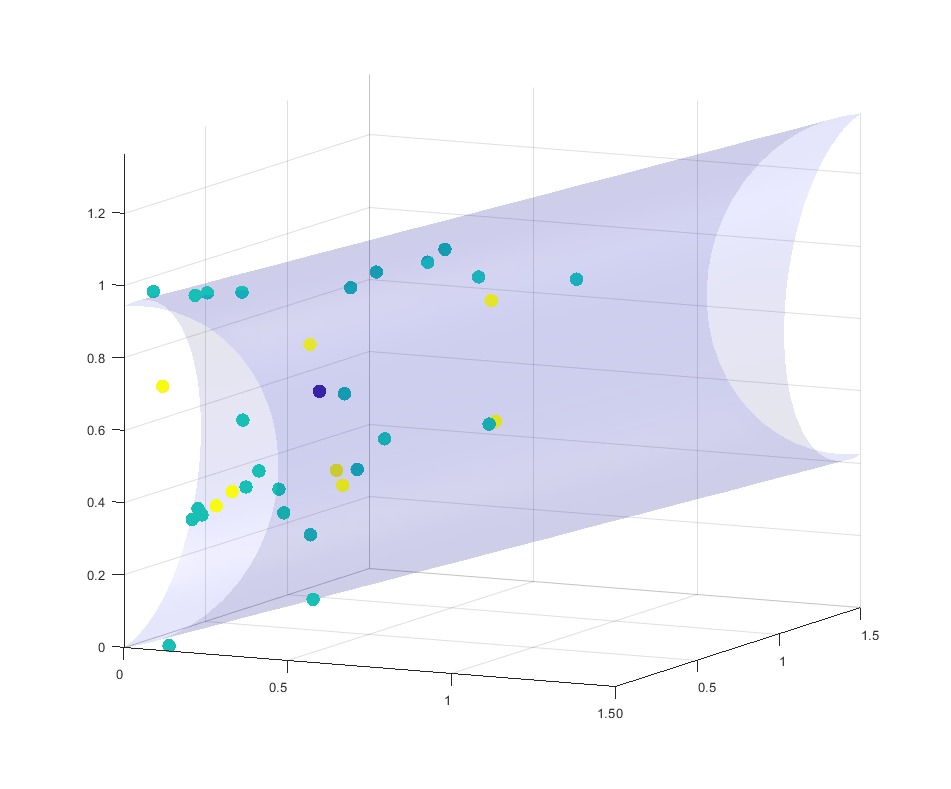}{(e)}
    \includegraphics[width=0.46\columnwidth]{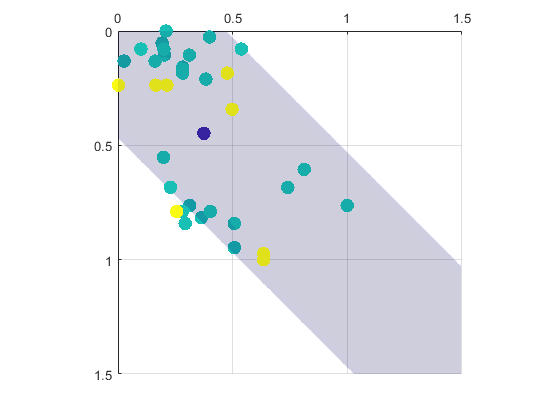}{(f)}
    
    \caption{Experiment result visualization diagram. The red, yellow and black ball in subplots (a) and (b) represent the states GUT=2, 1 and 0 respectively. The yellow, green and purple circle in remaining subplots represent the GUT=2, 1, 0 states. We obtained the tunnel in subplot (c), (d), (e) , and (f) by fitting the points in the experimental results where the user was in the good GUX state. Subplots (d) and (f) are the projections of subplot (c), (e) in the skill and challenge planes, respectively.}
\end{figure*}

\begin{table*}
  \caption{Analysis indicators of various methods.}
  \label{tab:commands}
  \begin{tabular}{ccccl}
    \toprule
Algorithm & ACC($\%$) & Precision($\%$) & Recall($\%$) & $F_{1}$-score($\%$) \\
    \midrule
Siamese network	& 43.22	& 45.97	& 50.00	& 41.75 \\
FC+PE&	69.49	& 57.04	& 47.30	& 48.07 \\
FC	 &55.93	&48.08	&46.40	& 42.67 \\
\textbf{GUT(Ours)} &	\textbf{76.27}	& \textbf{71.28}	&\textbf{75.41}	& \textbf{72.87}\\
    \bottomrule
  \end{tabular}
\end{table*}
 From the results in Table 3, our network classification results are better than the baseline methods. The accuracy and recall rates are higher than the other networks. 
 
 Mapping our experimental results to Motivation-Skill-Challenge 3D space, as illustrated in Figure 13. We can find that the distribution of the projection of our experimental results on the Skill-Challenge plane is basically in line with the distribution of Flow. The distribution of the experimental results in space is largely consistent with the shape of the tunnel, with most GUT=2 states being surrounded by GUT=1 states, which is largely consistent with the assumptions of our theoretical model.
 \subsection{GUT Visualization}
Visualisation tools address the challenge of analysing and presenting overwhelming amounts of data \cite{10.1145/2212776.2223795}. One of the challenges of GUX visualisation is making the interpretation of game process data (game event) and user psychology data (GUX) meaningful in terms of facilitating design decisions for developers. To help game designers quickly and conveniently located game events and understand user experiences, we propose two kinds of data visualisation schemes, as illustrated in Figure 14 and 15 which includes both the key of game process data and the users experience during the match. This helps game designers to understand what happens to players during the game play, and how the players feel in contexts of the game events. The plan of figure 14 can help game designers find the relationships between the user GUX states and team-fights events. The plan of figure 15 can help game designers find the key time in the game map, and optimize the design of maps. At the same time, we make a user GUX visualization system based on the result from GUXAS tool, as illustrated in Figure 16. It includes both of the two kinds of data visualisation schemes, which can show the users' in-game GUX state and game process data to the game designers.

\begin{figure*}
  \centering
 \includegraphics[width = 1\textwidth]{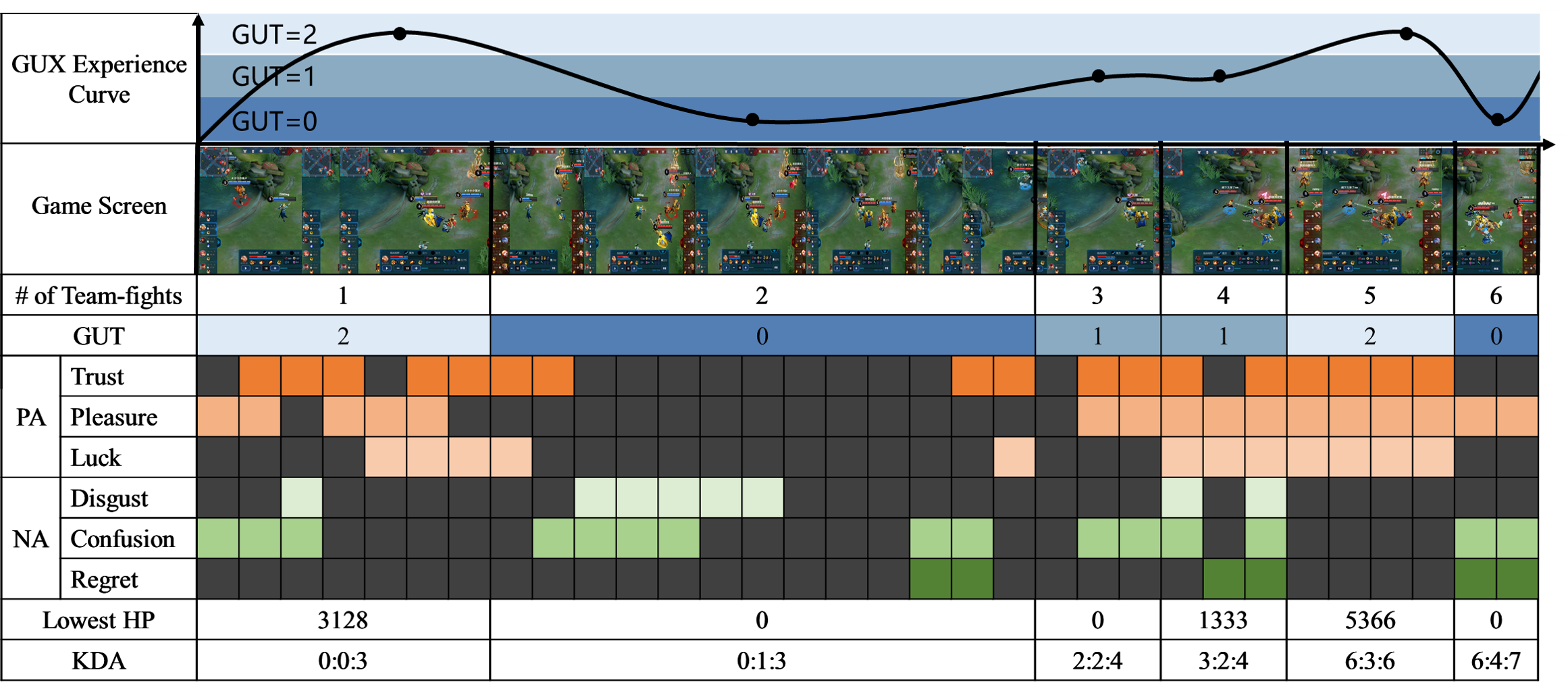}
  \caption{
 Data visualisation includes user group battle screenshots, GUX experience curves, player minimum blood, player KDA (kills:deaths:assists) and player GUX states in the team-fights. The Dark gray indicates that the affect is not present. The circles on the curve are the player's user experience level at that moment of the game. In the heat-map and the GUX Experience Curve, the dark blue, blue, and light blue represent GUT=2, GUT=1 and GUT=0 respectively.
}

\end{figure*}
   \begin{figure*}
  \centering
 \includegraphics[width = \textwidth]{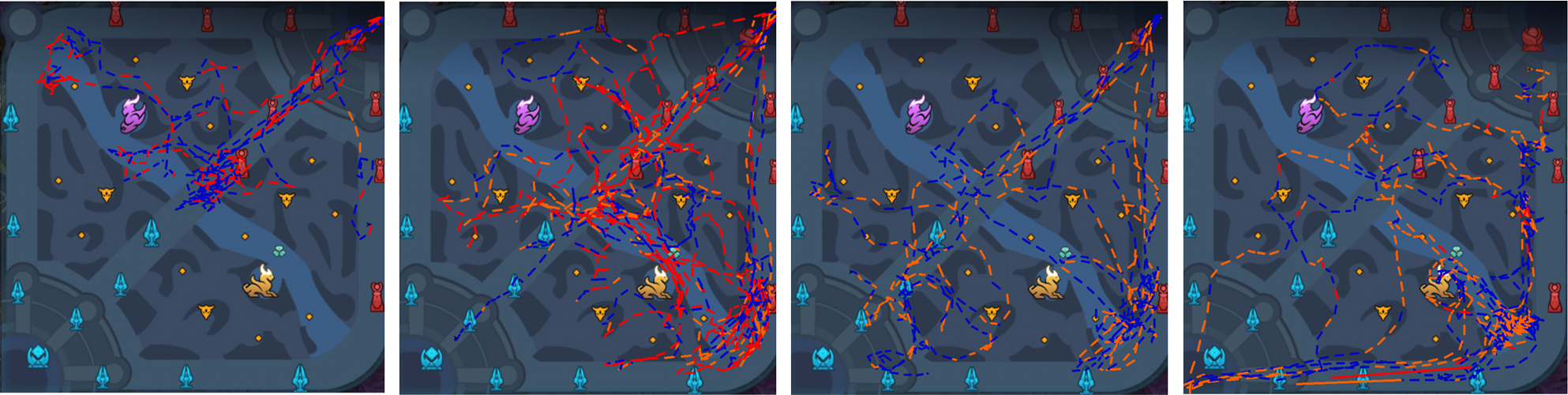}
  \caption{
The curves in the image represent the player's movement trajectory within the game. The orange, red, and blue colors represent the player's GUT=1, GUT=2, and GUT=0 states, respectively.}
\end{figure*}

\begin{figure*}
  \centering
 \includegraphics[width = 1\textwidth]{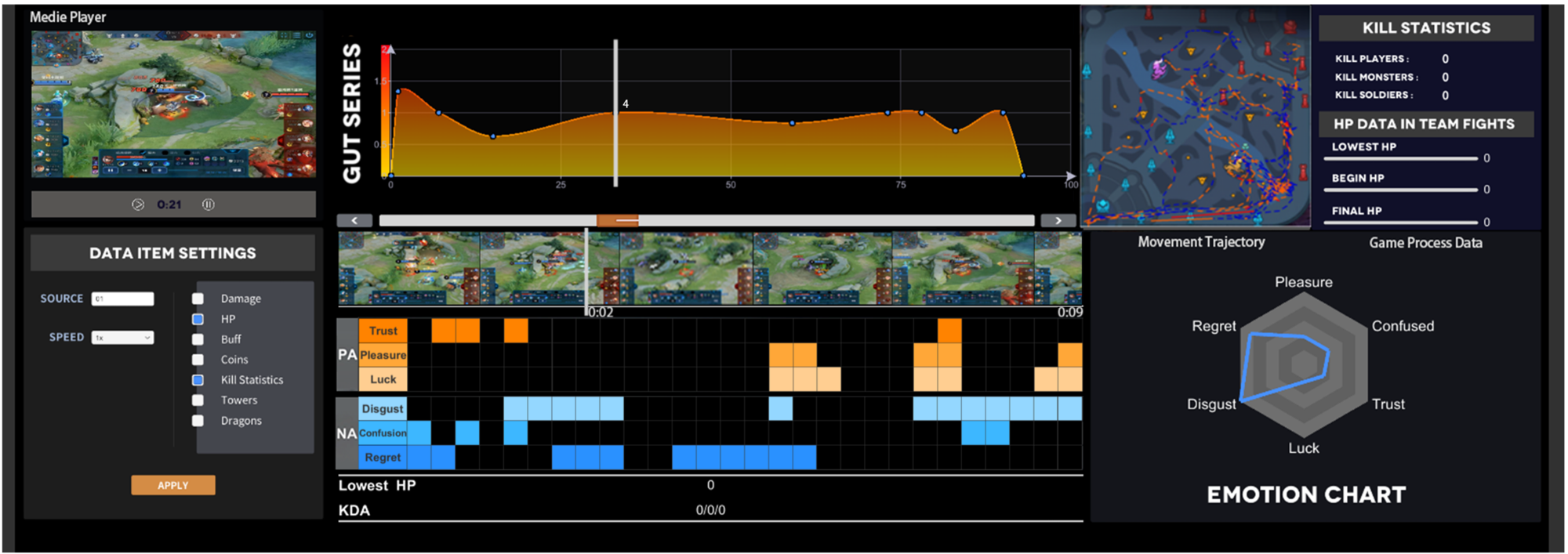}
  \caption{
Visualization system diagram. The system has the ability that can display the game video, showing the GUX experience curve, player's movement trajectory, game process data, affect heat-map, and affect statistical charts. It also has a system control panel.
}
\end{figure*}

\section{Discussion}
According to the result obtained from the physiological data and game process data, we visualize the data in the visualization system as illustrated in Figure 14. We combined the game process data with GUX state label, and found some phenomenon which might help the game designers better understand the player's GUX during the in-game session.

The first phenomenon noticed is that players' GUX state fluctuate strongly before and after purchasing a "gear". In other words, there are strong game incentives that affect players' GUX state at this time.
"Gears" are in-game items that a hero can purchase. They provide bonus attributes with unique abilities. 

In addition, on the basis of first phenomenon (\\phenomenon 1), we also found phenomenon 2. On the whole, players GUX state change process can be categorized based on the purchase of gears.
We can partition the MOBA timeline as 'fluctuation' → good GUX state → GUT=0 state (the player is not well immersed in the game world), 
corresponds to Purchase Gears→ Combat → Farming  in the game stage.

Among them, the 'fluctuation' stage refers to the player's GUX state has a higher frequency swing from GUT=2, \\GUT=1 to GUT=0. Which means that the player's GUX state is in an unstable state at this time.
Purchase Gears refers to the behavior of players changing accumulated game gold into gears in the game, which means players will update hero attributes with the accumulated game economic values. 
Combat means that the player has just updated the hero attributes, and the player will be very excited to verify his newly updated attributes by team-fights or killing the game mechanism. The player is attracted to the game at this stage.  
Farming is the act of killing waves of Minion and Monster to gather Gold and Experience. During this time the players' skill is much higher than the challenge he has to face. At the same time, he has to cultivate (farm) to purchase new gears to  win the match, which will make the game boring to players.

In addition to the above two phenomena, relying on the experimental results, we also found the following phenomena. The good GUX (GUT=1,2) occurs if the player kills the in-game mechanic (like Tyrant, Master) or after a team-fights occurs and the player does not die during the team-fights time. GUT=0 status occurs when the player is killed or injured during the game. 



\section{Conclusion}

In this paper, we proposed a new GUX evaluation method named GUT. Using the GUT model and GUXAS tool, we can assign GUX labels to game process data using physiological data. The method can also predict the player’s GUX state using game process data (game log). We expand the existing work and study a more general GUX detection model using the physiological data and game process data.
Since there is no related works in our research work, there are still some limitations in the verification of the model. We studied only one kind of game and thus some of our results might be specific to that. We also had a relatively limited number of participants (53). In future work, we will continuously optimize our GUX Field evaluation method, and propose a more convenient and general GUX accuracy evaluation method. At the same time, we will try to use our model and tool in other kinds of games like First-person shooting (FPS) and Real-Time Strategy (RTS) Games. We will make our tool more convenient to use to help game designers develop intriguing games and also promote game discipline.

\bibliographystyle{ACM-Reference-Format}
\bibliography{sample-base}










\end{document}